\documentclass[aps,pra,onecolumn,showpacs]{revtex4}
\usepackage{amsmath,epsfig,amsfonts,graphicx}
\usepackage{subfigure}
\usepackage{amssymb,amsthm,newlfont,setspace,rotating}
\usepackage{color}
\usepackage[normalem]{ulem}

\newcommand{\be}{\begin{equation}}
\newcommand{\ee}{\end{equation}}

\begin{document}

\title{Unstaggered-staggered solitons in two-component discrete nonlinear
Schr\"{o}dinger lattices}
\author{Boris A. Malomed$^{1}$, D. J. Kaup$^{2}$, and Van~Gorder, Robert A.$%
^{3}$}

\begin{abstract}
We present stable bright solitons built of coupled unstaggered and
staggered components in a symmetric system of two discrete nonlinear
Schr\"{o}dinger (DNLS) equations with the attractive
self-phase-modulation (SPM)\ nonlinearity, coupled by the repulsive
cross-phase-modulation (XPM) interaction. These mixed modes are of a
``symbiotic" type, as each component in isolation may only carry
ordinary unstaggered solitons. The results are obtained in an
analytical form, using the variational \ and Thomas-Fermi
approximations (VA and TFA), and the generalized Vakhitov-Kolokolov
(VK) criterion for the evaluation of the stability. The analytical
predictions are verified against numerical results. Almost all the
symbiotic solitons are predicted by the VA quite accurately, and are
stable. Close to a boundary of the existence region of the solitons
(which may feature several connected branches), there are broad
solitons which are not well approximated by the VA, and are
unstable.
\end{abstract}

 %03.75.Lm; 03.75Nt; 

\address{$^{1}$Department of Physical Electronics, School of Electrical
Engineering, Faculty of Engineering, Tel Aviv University, Tel Aviv
69978,
Israel\\
$^{2}$Department of Mathematics \& Institute for Simulation and Training\\
University of Central Florida, Orlando, FL 32816-1364 USA\\
$^{3}$Department of Mathematics, University of Central Florida, Orlando, FL 32816-1364 USA\\
}

\maketitle

%\noindent \textit{Keywords}: discrete nonlinear Schr\"{o}dinger equation;
%soliton; stability

\section{Introduction}

Discrete nonlinear Schr\"{o}dinger (DNLS) equations constitute a class of
lattice models which comprise diverse physical settings \cite{Panos}. A
straightforward realization of the DNLS equation in arrays of evanescently
coupled optical waveguides was first proposed in Ref. \cite{Demetri}, and
later demonstrated experimentally in a set of parallel semiconductor
waveguides \cite{Silberberg}. Multi-core nonlinear waveguiding systems have
also been created in the form of optically-written virtual lattices in
photorefractive materials~\cite{Moti}, and as permanent structures written
by laser pulses in bulk silica \cite{Jena}. A thorough review of the
nonlinear discrete optics, developed experimentally and theoretically in
these and allied media, was given in Ref. \cite{review}. The DNLS equations
find another important application in modeling the mean-field dynamics of
Bose-Einstein condensates (BECs)\ loaded into deep optical-lattice
potentials. In this case, it was demonstrated experimentally \cite%
{discreteBECexperiment} and theoretically \cite{discreteBECtheory} that the
periodic potential effectively splits the condensate into a set of droplets
trapped in local potential wells, which are linearly coupled by tunneling of
atoms across the separating potential barriers, DNLS equations being natural
models for such quasi-discrete systems.

Two fundamental types of discrete solitons supported by the DNLS equations
with the self-repulsive and self-attractive on-site nonlinearity are
localized modes of \textit{staggered} and \textit{unstaggered} types,
respectively, i.e., ones with opposite signs of the lattice field at
adjacent sites, or without the sign alternation \cite{Panos}. In the
continuum limit, the unstaggered solitons correspond to regular ones,
residing in the semi-infinite gap of the continual NLS equation, while the
staggered solitons may be considered as counterparts of gap solitons, which
exist in finite bandgaps of the spectrum induced by a periodic potential, in
the case of the self-defocusing nonlinearity \cite{gapsol}.

A natural generalization, which also finds many applications to optics and
BEC, is represented by systems of coupled DNLS equations. In optics, the
system models the co-propagation of two waves carried by different
polarizations or wavelengths in the same waveguiding array, while in BEC the
coupled equations describe a mixture of two condensates, which may represent
either different hyperfine states of the same atomic species, or two
different kinds of atoms \cite{Pit}. Normally, two-component discrete
solitons in such systems feature one type of the intrinsic structure,
unstaggered or staggered, in both components, because the signs of the
self-phase-modulation (SPM) nonlinearity acting on each component, and of
the cross-phase-modulation (XPM) nonlinearity which couples them, are the
same \cite{Panos}. The objective of the present work is to introduce
two-component discrete solitons of the \textit{mixed} type, built as
combinations of unstaggered and staggered components. Previously,
single-component surface modes of a mixed unstaggered-staggered type were
studied at an interface between two different lattices \cite{Molina}, but,
to the best of our knowledge, two-component mixed solitons were not reported
before. On the other hand, in bimodal continual systems with the periodic
potential acting on both components, solitons of a \textit{semi-gap} type,
which may be considered as continuous counterparts of the discrete ones
introduced in the present work, were studied in Ref. \cite{semi}. They are
composed of an ordinary soliton in one component and a gap soliton in the
other. The semi-gap solitons are somewhat similar to the earlier studied
\textit{intergap} solitons, that were built as bound states of two
components represented by solitons belonging to two different finite
bandgaps (the first and second ones) \cite{Arik}.

The paper is organized as follows. The model is formulated in Section II.
Approximate analytical results are presented in Section III. These results,
based chiefly on the variational approximation (VA), demonstrate that the
mixed unstaggered-staggered solitons are possible in the symmetric system of
DNLS\ equations when the XPM interaction between the two components is \emph{%
repulsive}, on the contrary to the self-attractive SPM nonlinearity. The
situation with the opposite signs of the SPM and XPM terms seems exotic in
optics, but it is quite possible in BEC, where the sign of the interactions
may be readily switched by means of the Feshbach resonance (see, e.g., Ref.
\cite{Feshbach}). In the case of strong difference between masses of the two
components, another analytical solution is obtained, based on the
Thomas-Fermi approximation (TFA). Numerical results, which allow us to
outline existence regions of fundamental (single-peak) solitons combining
the unstaggered and staggered components, and identify their stability
(almost all the solitons are stable), are summarized in Section IV.
Analytical results for the stability are reported too, based on the
Vakhitov-Kolokolov (VK) criterion for the two-component system. The
numerical results corroborate the predictions of the VA quite well; in
particular, it is confirmed that the mixed unstaggered-staggered solitons
exist only in the case of the repulsive XPM, whilst the SPM is
self-attractive in both components. The paper is concluded by Section V.

\section{The model}

The underlying system of the DNLS equations for lattice fields $\phi _{n}$
and $\psi _{n}$ is
\begin{subequations}
\label{phipsi}
\begin{eqnarray}
i\frac{d}{dt}\phi _{n} &=&-\frac{1}{2}\left( \phi _{n+1}+\phi _{n-1}-2\phi
_{n}\right) -\left( \left\vert \phi _{n}\right\vert ^{2}+\beta \left\vert
\psi _{n}\right\vert ^{2}\right) \phi _{n}, \\
i\frac{d}{dt}\psi _{n} &=&-\frac{1}{2m}\left( \psi _{n+1}+\psi _{n-1}-2\psi
_{n}\right) -\left( \left\vert \psi _{n}\right\vert ^{2}+\beta \left\vert
\phi _{n}\right\vert ^{2}\right) \psi _{n},
\end{eqnarray}%
where $t$ is time in the case the BEC mixture, or the propagation distance
in the array of optical waveguides, $m$ is the relative atomic mass of the
two species in the case of BEC, or the inverse ratio of the inter-site
coupling constants in the waveguide array, and $\beta $ is the relative
coefficient of the XPM coupling between the fields, assuming that the
coefficients of the self-attractive SPM nonlinearity for both fields are
scaled to be $1$. It should be mentioned that the model based on Eqs. (\ref%
{phipsi}) is not the most general one, as, rescaling both fields to make
their SPM coefficients equal to $1$, one can make the XPM interaction
asymmetric, with two different coefficients in Eqs. (\ref{phipsi}a) and (b),
$\beta _{\phi }\neq \beta _{\psi }$. Nevertheless, quite generic results
concerning the discrete solitons can be obtained within the framework of the
present system.

Solutions with unstaggered $\phi _{n}$ and staggered $\psi _{n}$ components
and two chemical potentials, $\lambda $ and $\mu $, are sought for as
\end{subequations}
\begin{equation}
\phi _{n}(t)=e^{-i\lambda t}u_{n},~\psi _{n}(t)=e^{-i\mu t}\left( -1\right)
^{n}v_{n},  \label{stag}
\end{equation}%
where real $u_{n}$ and $v_{n}$ satisfy the following stationary equations,
\begin{subequations}
\label{uv}
\begin{eqnarray}
\left( \lambda -1\right) u_{n}+\frac{1}{2}\left( u_{n+1}+u_{n-1}\right)
+\left( u_{n}^{2}+\beta v_{n}^{2}\right) u_{n} &=&0, \\
\left( \mu -\frac{1}{m}\right) v_{n}-\frac{1}{2m}\left(
v_{n+1}+v_{n-1}\right) +\left( v_{n}^{2}+\beta u_{n}^{2}\right) v_{n} &=&0,
\end{eqnarray}%
that can be derived from the\ Lagrangian,
\end{subequations}
\begin{eqnarray}
L &=&\frac{1}{2}\sum_{n=-\infty }^{+\infty }\left[ -\frac{1}{2}\left(
u_{n+1}-u_{n}\right) ^{2}+\lambda u_{n}^{2}+\frac{1}{2m}\left(
v_{n+1}-v_{n}\right) ^{2}+\left( \mu -\frac{2}{m}\right) v_{n}^{2}\right.
\notag \\
&&\left. +\frac{1}{2}u_{n}^{4}+\frac{1}{2}v_{n}^{4}+\beta u_{n}^{2}v_{n}^{2}%
\right] .  \label{Lagr}
\end{eqnarray}%
In the large-$m$ limit, which is tantamount to the TFA \cite{Pit} for
discrete equation (\ref{uv}b), this equation demonstrates that $v_{n}$ can
be eliminated in favor of $u_{n}$, hence in this case the coupled stationary
system reduces to a single equation.

In the next section, we present variational solutions based on an
exponential ansatz for fundamental (single-peak) solitons, and continue the
analysis in Section III by means of numerical methods. For given $\beta $
and $m$, we determine regions in the $(\lambda ,\mu )$ plane for which
single-peak numerical solutions exist and are stable. It is also found that
the related energy surfaces, i.e., norms of the two components as functions
of $\lambda $ and $\mu $, always decrease in $\lambda $ and either increase
or decrease monotonically in $\mu $, depending on the sign of $1-\beta ^{2}$%
. In this way, the generalized Vakhitov - Kolokolov (VK) stability criterion
for two-component solitary waves can be applied here \cite{VK,Berge'}. In
related two-component continuous systems \cite{CA, OKSF, YLP, YPV}, modeled
by coupled continual NLS equation, one can introduce a new parameter (the
ratio of $\lambda $ and $\mu $) and rescale the variables, to make the
stationary states depending on one (rather than two) effective chemical
potential \cite{YPV}. Moreover, a generalized VK stability criteria was
developed for a system of $N$ incoherently coupled continuous NLS equations
in Ref. \cite{PK}.

As for discrete systems, the single DNLS equation with the arbitrary
power-law nonlinearity was studied, by means of the variational
approximation (VA), in Ref. \cite{MW1996}, and the stability of
multi-soliton bound states in the DNLS equation with the cubic self-focusing
nonlinearity was investigated in Ref. \cite{stability}. A complex version of
the VA made it later possible to make predictions about collisions between
moving lattice solitons in the same basic model \cite{Papa}. Another
variational ansatz, relevant for DNLS solitons located on or anywhere
between lattice cites, was elaborated in Ref. \cite{DJK2005}. The VA was
further generalized for the DNLS equation with the cubic-quintic on-site
nonlinearity \cite{Ricardo}. Very recently, the accuracy of the VA-based
description of static discrete solitons and their stability, based on \emph{%
ans\"{a}tze }with different numbers of free parameters, was investigated in
a rigorous form in Ref. \cite{new}. As concerns discrete two-component
systems, the VA was used for studying the spontaneous symmetry breaking in
parallel DNLS lattices, linearly coupled at all sites \cite{Herring}, or at
a single site \cite{Sandra}.

\section{Analytical approximations}

\subsection{The variational approximation for the discrete solitons}

To apply the VA to the solution of Eqs. (\ref{uv}), we employ the
exponential ansatz that was earlier used in the framework of other models
\cite{MW1996}, \cite{DJK2005}, \cite{Ricardo}-\cite{Sandra}:
\begin{equation}
u_{n}=Ae^{-p|n|},~v_{n}=Be^{-q|n|}.  \label{ansatz}
\end{equation}%
We find the decay rates of the wave forms in Eq. (\ref{ansatz}), $p$ and $q$%
, not from the variational principle, but by requiring the ansatz to satisfy
the linearized limit of Eqs. (\ref{uv}) at $n\rightarrow \pm \infty $:
\begin{subequations}
\label{pq}
\begin{eqnarray}
p &=&\ln \left( 1-\lambda +\sqrt{-\lambda \left( 2-\lambda \right) }\right)
,~ \\
q &=&\ln \left( m\mu -1+\sqrt{m\mu \left( m\mu -2\right) }\right) \,.
\end{eqnarray}%
For $p$ and $q$ to be real and positive, the allowed ranges of chemical
potentials $\mu $ and $\lambda $ are
\end{subequations}
\begin{equation}
\lambda <0\,,\quad \mu -\frac{2}{m}>0\,.  \label{><}
\end{equation}

Substituting ansatz (\ref{ansatz}) into Lagrangian (\ref{Lagr}) and carrying
out the summation yields the effective Lagrangian,%
\begin{eqnarray}
2L_{\mathrm{eff}} &=&-A^{2}\tanh \left( \frac{p}{2}\right) +\frac{B^{2}}{m}%
\tanh \left( \frac{q}{2}\right) +\lambda A^{2}\coth p+\left( \mu -\frac{2}{m}%
\right) B^{2}\coth q  \notag \\
&&+\frac{A^{4}}{2}\coth \left( 2p\right) +\frac{B^{4}}{2}\coth \left(
2q\right) +\beta A^{2}B^{2}\coth \left( p+q\right) ,  \label{Leff}
\end{eqnarray}%
which gives rise to the variational equations, $\partial L_{\mathrm{eff}%
}/\partial \left( A^{2}\right) =\partial L_{\mathrm{eff}}/\partial \left(
B^{2}\right) =0$, i.e.,
\begin{subequations}
\label{eqs}
\begin{eqnarray}
A^{2}\coth \left( 2p\right) +\left[ \beta \coth \left( p+q\right) \right]
B^{2} &=&\tanh \left( \frac{p}{2}\right) -\lambda \coth p, \\
\left[ \beta \coth \left( p+q\right) \right] A^{2}+B^{2}\coth \left(
2q\right)  &=&-\frac{1}{m}\tanh \left( \frac{q}{2}\right) -\left( \mu -\frac{%
2}{m}\right) \coth q~.
\end{eqnarray}%
As seen from Eq. (\ref{eqs}b) and (\ref{><}), solutions to the variational
equations with positive $A^{2}$ and $B^{2}$ do not exist in the case of $%
\beta >0$, but a solution \emph{may exist} at $\beta <0$.

The fact that the fundamental solitons of the mixed unstaggered-staggered
type may exist as the bound state of two components, which, in isolation,
support solely ordinary unstaggered solitons (through the self-attractive
SPM), suggests to identify the solitons of the mixed type as \textit{%
symbiotic} ones, cf. Ref. \cite{symbio}, where symbiotic solitons were
defined in the opposite case, for the continual system with the
self-repulsive SPM and attractive XPM nonlinearities. On the other hand, the
staggering effectively reverses the signs of the SPM nonlinearity and
external potential, therefore, in the presence of a large-amplitude
unstaggered component, the staggered one may be considered as a soliton with
the intrinsic self-repulsive nonlinearity, trapped in the attractive
external potential. Such a mode tends to exist and be stable, unless the
effective intrinsic self-repulsion is too strong, making the existence of
the trapped mode impossible \cite{Carr}.

We also note (this remark will be relevant for comparison with some
numerical results presented in the next section) that a solution to Eqs. (%
\ref{eqs}), considered as a linear system for $A^{2}$ and $B^{2}$, may not
exist when the determinant of the system vanishes, i.e.,
\end{subequations}
\begin{equation}
\coth (2p)\coth (2q)-\beta ^{2}\coth ^{2}(p+q)\,=0.  \label{det}
\end{equation}%
Nevertheless, a solution is possible under condition (\ref{det}) if the
right-hand sides of Eqs. (\ref{eqs}) are related in the same way as the two
rows of the degenerate determinant, i.e.,
\begin{equation}
\tanh \left( \frac{p}{2}\right) -\lambda \coth p=-\frac{\coth (2p)}{\beta
\coth (p+q)}\left[ \frac{1}{m}\tanh \left( \frac{q}{2}\right) +\left( \mu -%
\frac{2}{m}\right) \coth q\right] \,.  \label{consistent}
\end{equation}

\subsection{Three-layer solitons for $1/m\rightarrow 0$ (the discrete
Thomas-Fermi approximation)}

There is another case in which we can determine properties of the solution
in an analytical form. When the staggered species is very heavy, i.e., $%
m\rightarrow \infty $ in Eq. (\ref{phipsi}b), the second equation from
system (\ref{uv}), at lowest order, takes the local form:
\begin{equation}
\left[ \mu +\left( v_{n}^{2}+\beta u_{n}^{2}\right) \right] v_{n}=0.
\label{degen}
\end{equation}%
Equation (\ref{degen}) has three possible solutions, \textit{viz}.%
\begin{equation}
v_{n}^{2}=-\mu -\beta u_{n}^{2}\,,  \label{v^2}
\end{equation}%
or $v_{n}=0$, which may be used to eliminate $v_{n}$ in favor of $u_{n}$,
cf. a similar approach allowing one to eliminate a heavy fermionic component
in \ Bose-Fermi mixture \cite{Padova}. Accordingly, discrete solitons,
composed of three layers, can be built as follows: in the central region (%
\textit{inner layer}), we use relation (\ref{v^2}) and substitute it into
the first equation of system (\ref{uv}), which yields%
\begin{equation}
\left[ \left( \lambda -1\right) -\beta \mu \right] u_{n}+\frac{1}{2}\left(
u_{n+1}+u_{n-1}\right) +\left( 1-\beta ^{2}\right) u_{n}^{3}=0,  \label{u}
\end{equation}%
i.e., the stationary DNLS equation which gives rise to soliton solutions.
Requiring this solution, in the central region, to be a part of a discrete
soliton with a single peak and centered at $n=0$, then one must have $\beta
^{2}<1$ and $\lambda <\beta \mu $.

It follows from Eq. \eqref{v^2} that, since $\mu >0$ [see Eq.
(\ref{><})], for $v_{n}^{2}$ to be positive, one must take $-1<\beta
<0$, which yields
%With regard to the above conditions $\mu >0$ [see Eq. (\ref{><})] and $\beta
%^{2}<1$, Eq. \eqref{v^2} may yield physically relevant solutions, with $%
%v_{n}^{2}>0$, only in the case of $-1<\beta <0$:
$v_{n}^{2}=\left( -\beta \right) \left( u_{n}^{2}-U^{2}\right) $, with $%
U^{2}\equiv -\mu /\beta >0$. Provided that $u_{0}^{2}>U^{2}$, one thus has $%
v_{n}^{2}>0$ at $n=0$ and in some region around $n=0$ (in the \textit{inner
layer} of the solution, as defined above), However, the positiveness of the
so defined $v_{n}^{2}$ will be lost at $|n|>N$ with $N$ large enough, as $%
u_{n}^{2}$ for soliton solutions decays at $n\rightarrow \infty $. Thus, for
$n>N$ and $n<-N$ (in the two outer layers), the discrete mode can be
extended upon taking the other root of Eq. (\ref{degen}) for $v_{n}$,
namely, $v_{n}\equiv 0$, which thus causes $u_{n}$ to satisfy the usual DNLS
equation, following from Eq. (\ref{uv}b) with $v_{n}\equiv 0$:%
\begin{equation}
\left( \lambda -1\right) u_{n}+\frac{1}{2}\left( u_{n+1}+u_{n-1}\right)
+u_{n}^{3}=0.  \label{NLS}
\end{equation}%
Obviously, Eq. (\ref{NLS}) has usual solution vanishing as $|n|\rightarrow
\infty $ for $\lambda <0$ [recall $\lambda <0$ is imposed by Eq. (\ref{><}%
)], thus the composite soliton can be constructed by combining the
appropriate solutions in the inner and outer layers. The conditions of
matching the discrete fields at $n=\pm N$, which includes setting $v_{N}=0$
(as required by the TFA), imposes two constraints on the set of parameters $%
\lambda $, $\mu $, and $N$, hence the solution is expected to exist along a
curve in the plane of $\left( \lambda ,\mu \right) $, which is corroborated
by numerical findings presented in the next section. Note that, in the
framework of the present approximation, there is actually \emph{no difference%
} between the unstaggered and staggered forms of the solution for $v_{n}$,
as only $v_{n}^{2}$ is determined by Eq. (\ref{v^2}).

\section{Numerical soliton solutions}

\subsection{The formulation of the numerical problem}

We look for numerical solutions to Eqs. \eqref{uv} for spatially
symmetric solitons, with $u_{-n}=u_{+n}$, $v_{-n}=v_{+n}$, and both
fields $u_{n}$ and $v_{n}$ monotonously decaying with the increase
of $n$, but never changing their signs, to support the unstaggered
and staggered shapes of the underlying components $\phi _{n}$ and
$\psi _{n}$, respectively, according to Eq. (\ref{stag}). At $n=0$,
Eqs. \eqref{uv} yield
\begin{subequations}
\label{11conds}
\begin{eqnarray}
u_{1} &=&-\left[ u_{0}^{2}+\beta v_{0}^{2}+\left( \lambda -1\right) \right]
u_{0}, \\
v_{1} &=&m\left[ v_{0}^{2}+\beta u_{0}^{2}+\left( \mu -\frac{1}{m}\right) %
\right] v_{0}.
\end{eqnarray}%
According to the above conditions, solutions to Eqs. (\ref{11conds}) must
satisfy constraints $u_{0}>u_{1}>0$ and $v_{0}>v_{1}>0$, thereby implying
that
\end{subequations}
\begin{subequations}
\label{11ineq}
\begin{eqnarray}
-1 &<&u_{0}^{2}+\beta v_{0}^{2}+\left( \lambda -1\right) <0, \\
\frac{1}{m} &<&v_{0}^{2}+\beta u_{0}^{2}+\mu <\frac{2}{m}\,.
\end{eqnarray}%
Continuing in this manner, i.e., imposing bounds $u_{1}>u_{2}>0$, $%
v_{1}>v_{2}>0$, and so on, as it follows from Eqs. \eqref{uv} at
$n=1,2,$ ..., one successively restricts the region of the $(\lambda
,\mu )$ plane in which the soliton solutions are possible.

The numerical solution of Eqs. \eqref{uv} was carried out by means
of a discrete version of the shooting method, which used the
VA-predicted solution as the initial guess, and was iterated until
discrete wave forms monotonously decreasing with $n$ without the
change of the sign, up to the level of $\left( u_{n},v_{n}\right)
\sim 10^{-5}\left( u_{0},v_{0}\right) $, were found. 

For stability
testing, we introduced initial perturbations, multiplying the
stationary solutions by
\end{subequations}
\begin{equation}
\lbrack 1+\delta \,\exp (ikn)],  \label{k}
\end{equation}%
with perturbation amplitude $\delta \simeq 5\%$ and $k\,N\sim 1$,
where $N$ is the effective size of the discrete soliton. Then, the
evolution of the thusly perturbed solution was simulated forward in time
until $t=50$. The results of the simulations were characterized by
``stability numbers" $S_{\phi }$ and $S_{\psi }$ for the two
components, which are defined as root-mean-square changes in the
\textit{relative} amplitude of the solution, compared to the initial
values, over the part of the lattice
where the discrete soliton is located. For stable solutions, we obtain $%
\left\vert S_{\phi },S_{\psi }\right\vert \ll 1$, while for unstable ones $%
\left\vert S_{\phi },S_{\psi }\right\vert $ grow to values $\gtrsim 1$.

Since these solitons are symbiotic, one might suspect that they could be
unstable to efforts to pull their two components apart. We have also checked
for this possibility numerically, as above, by taking wavenumbers $k_{u}$
and $k_{v}$ with opposite signs in perturbation factors (\ref{k}) for the
two fields. All solutions that we tested in this way, which had tested out
to be stable against other perturbations, were found to be stable in this
sense too.

%\textbf{Please write at least an outline of the numerical method here.
%Include both the method of obtaining the numerical solutions and then
%testing for stability from the time evolution. I am calling the stability
%numbers $S_\phi$ and $S_\psi$.}

\subsection{Dependence of solutions on the parameters}

In agreement with the prediction of the VA, numerical solutions for the
solitons were found solely for $\beta <0$, and, as suggested by Eq. (\ref{u}%
), $\beta =-1$ is a critical value. When $m=1$, we can demonstrate this 
with numerical results which makes
it possible to identify two distinct cases, $\beta <-1$ and $-1<\beta <0$,
seen in Fig. 1. When $\beta $ approaches $-1$ from either side, we find
solutions in a region which shrinks toward the line
\begin{equation}
\mu =2-\lambda   \label{ml}
\end{equation}%
in the $(\lambda ,\mu )$-plane. It is worthy to note that, as can be
found from inspection of Eqs. (\ref{eqs}) and (\ref{consistent}),
both these equations reduce precisely to Eq. (\ref{ml}) in the case
of $m=1$ and $\beta =-1$, i.e., only the ``double-degenerate"
solution selected by Eqs. (\ref{eqs}) and (\ref{consistent})
survives in this case. Note also that Eqs. (\ref{pq}) with $m=1$
yield equal decay rates $p$ and $q$ for the two components of the
soliton exactly under condition Eq. (\ref{ml}), i.e., the soliton
surviving in the limit of $m=1$ and $\beta =-1$ is characterized by
equal localization lengths of the two components.
\begin{figure}[tbp]
\begin{center}
\includegraphics[scale=0.25]{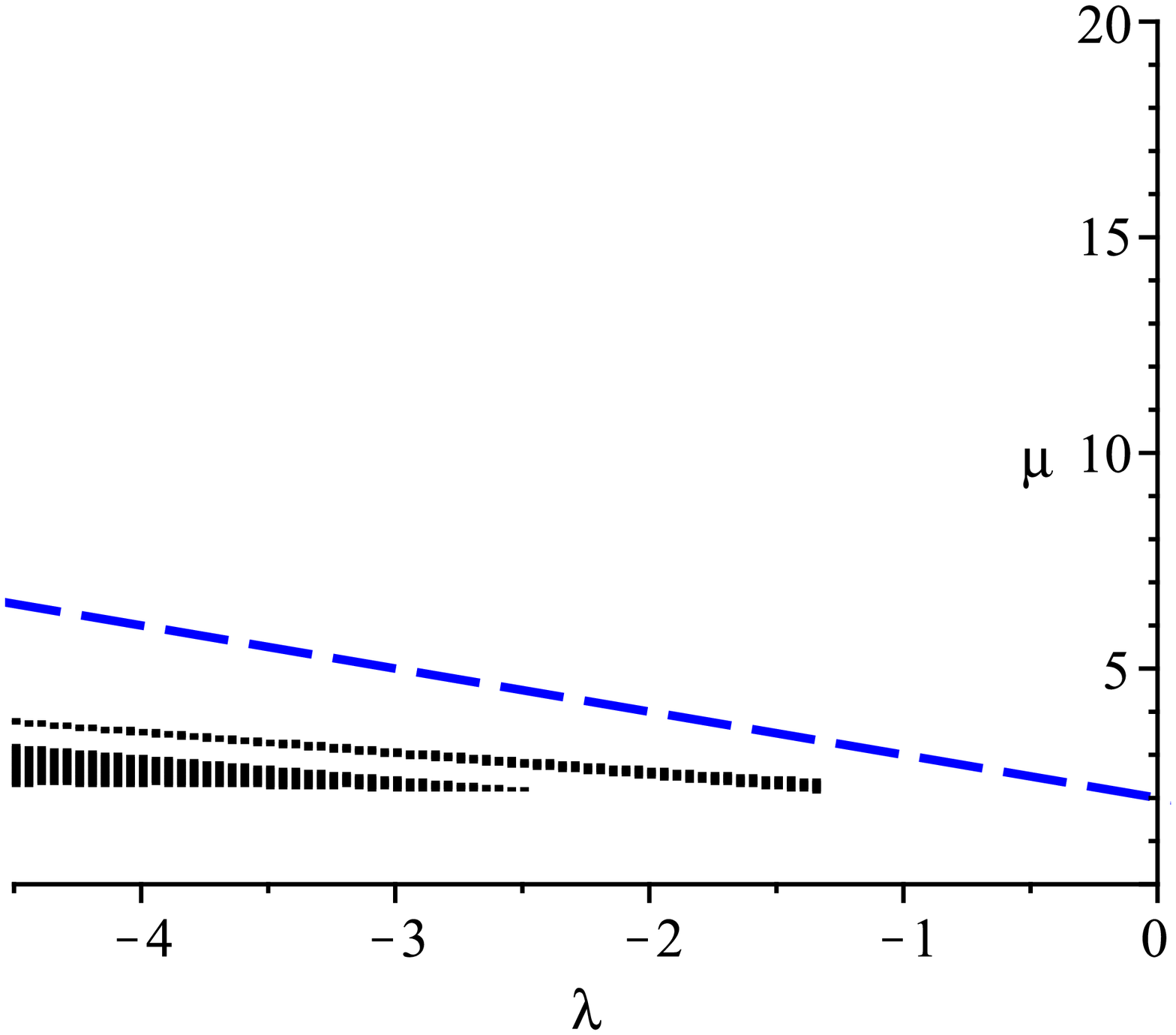} \includegraphics[scale=0.25]{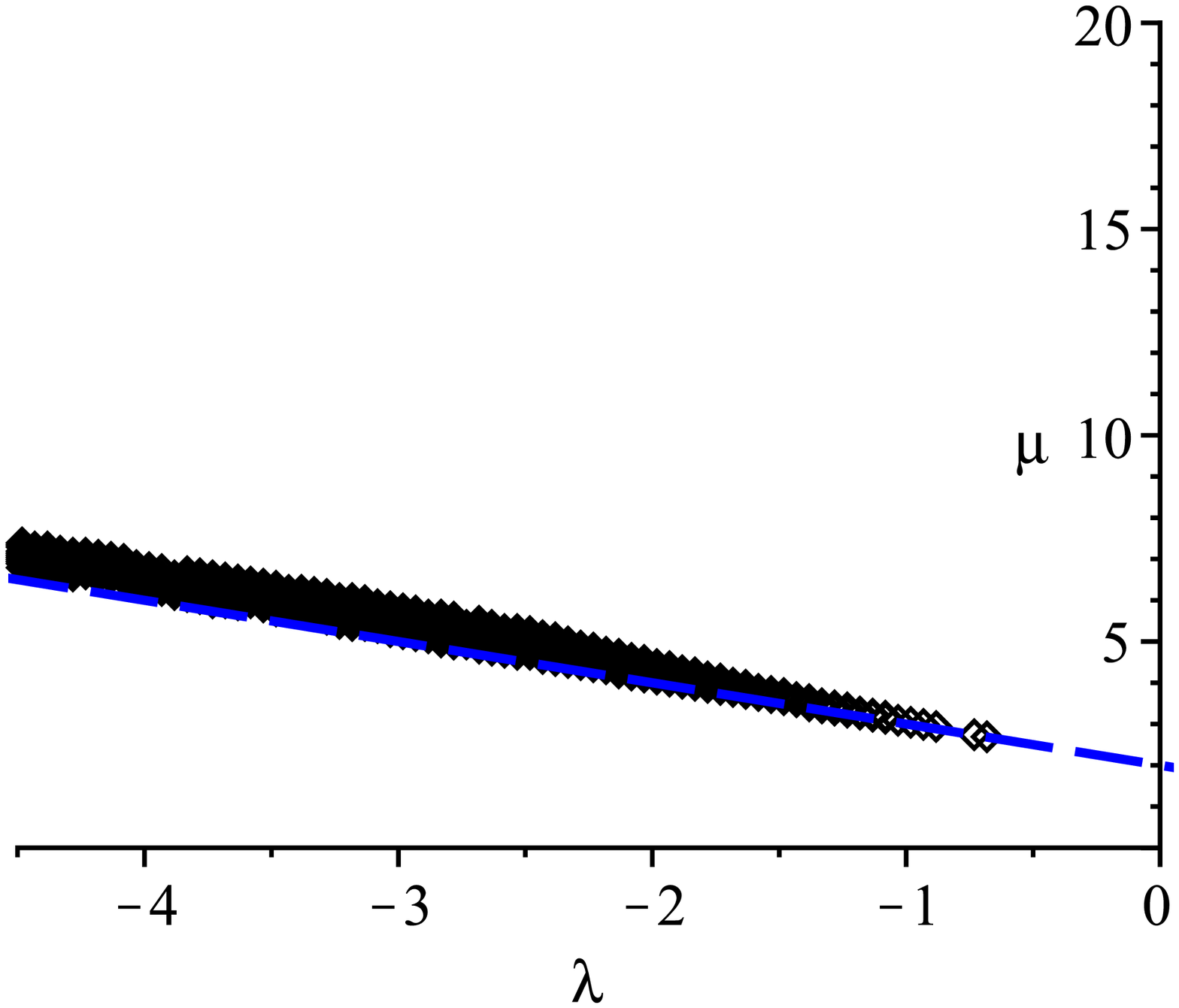} %
\includegraphics[scale=0.25]{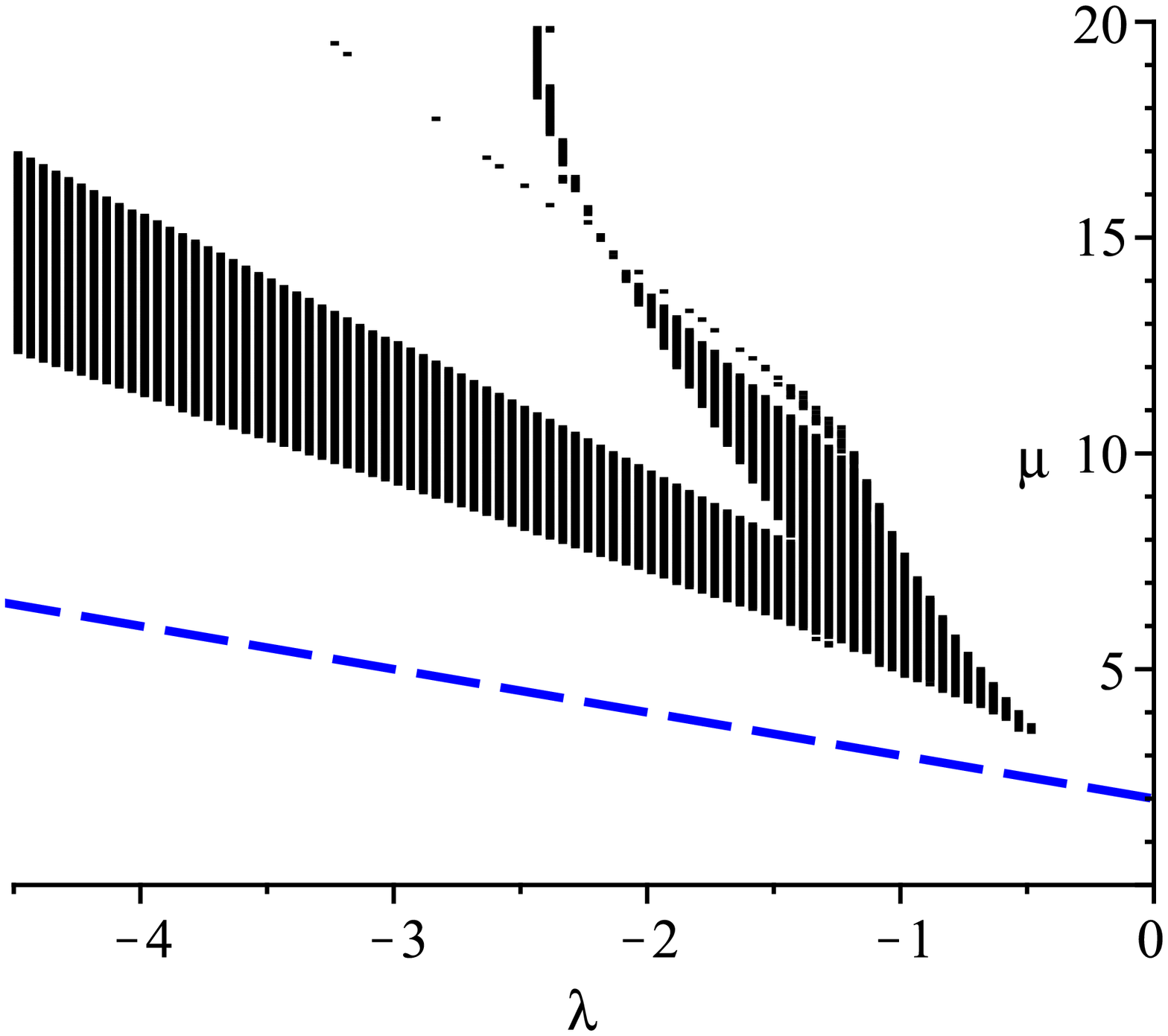} \vspace{-0.2in}
\begin{equation}
\text{(a)} \qquad \qquad \qquad \qquad \qquad \qquad \text{(b)} \qquad
\qquad \qquad \qquad \qquad \qquad \text{(c)}\notag
\end{equation}
\vspace{-0.5in}
\end{center}
\caption{(Color online) Existence regions (black) for the numerically found
single-peak discrete solitons in the $(\protect\lambda ,\protect\mu )$
plane, for $m=1$ and (a) $\protect\beta =-0.5$, (b) $\protect\beta =-1.01$,
(c) $\protect\beta =-2$. The dashed blue line corresponds to $\protect\mu =2-%
\protect\lambda $ [see Eq. (\protect\ref{ml})], to which the existence
region shrinks in the case of $\protect\beta =-1$.}
\end{figure}

Moving away from the critical value, $\beta =-1$, in either direction, Fig.
1 shows that the existence region of the numerically found solitons widens,
and, simultaneously, the region moves away from line (\ref{ml}), staying on
one side of this line, depending on the sign of $1+\beta $. For $\beta $
sufficiently far from $-1$, additional solution regions begin to split off
from the primary one in the $(\lambda ,\mu )$ plane. Additional solution
branches break off from the primary ``trunk" at small $%
|\lambda |$, then shrinking and disappearing as $|\lambda |$ increases,
while the main trunk widens as $\lambda \rightarrow -\infty $.

With the increase of the relative-mass parameter, $m$, the existence
region of the soliton solutions in the $(\lambda ,\mu )$ plane
shrinks, following Eqs. \eqref{11ineq}. This trend is observed in
Fig. 2, which suggests that
the region contracts toward a line in the $(\lambda ,\mu )$ plane at $%
m\rightarrow \infty $ (as predicted by the TFA presented above). In case $%
\beta $ is far enough from $-1$ to permit additional branches in the
existence diagrams, we observe that such branches collapse into the
primary one (the ``trunk"), which then itself collapses into a line,
as can be seen in Fig. 2 for $\beta =-2$. It is also worthy to note
that the bottom boundary of the existence region in Fig. 2 moves
upward with the increase of $m$ at fixed $\lambda $.
\begin{figure}[tbp]
\begin{center}
\includegraphics[scale=0.25]{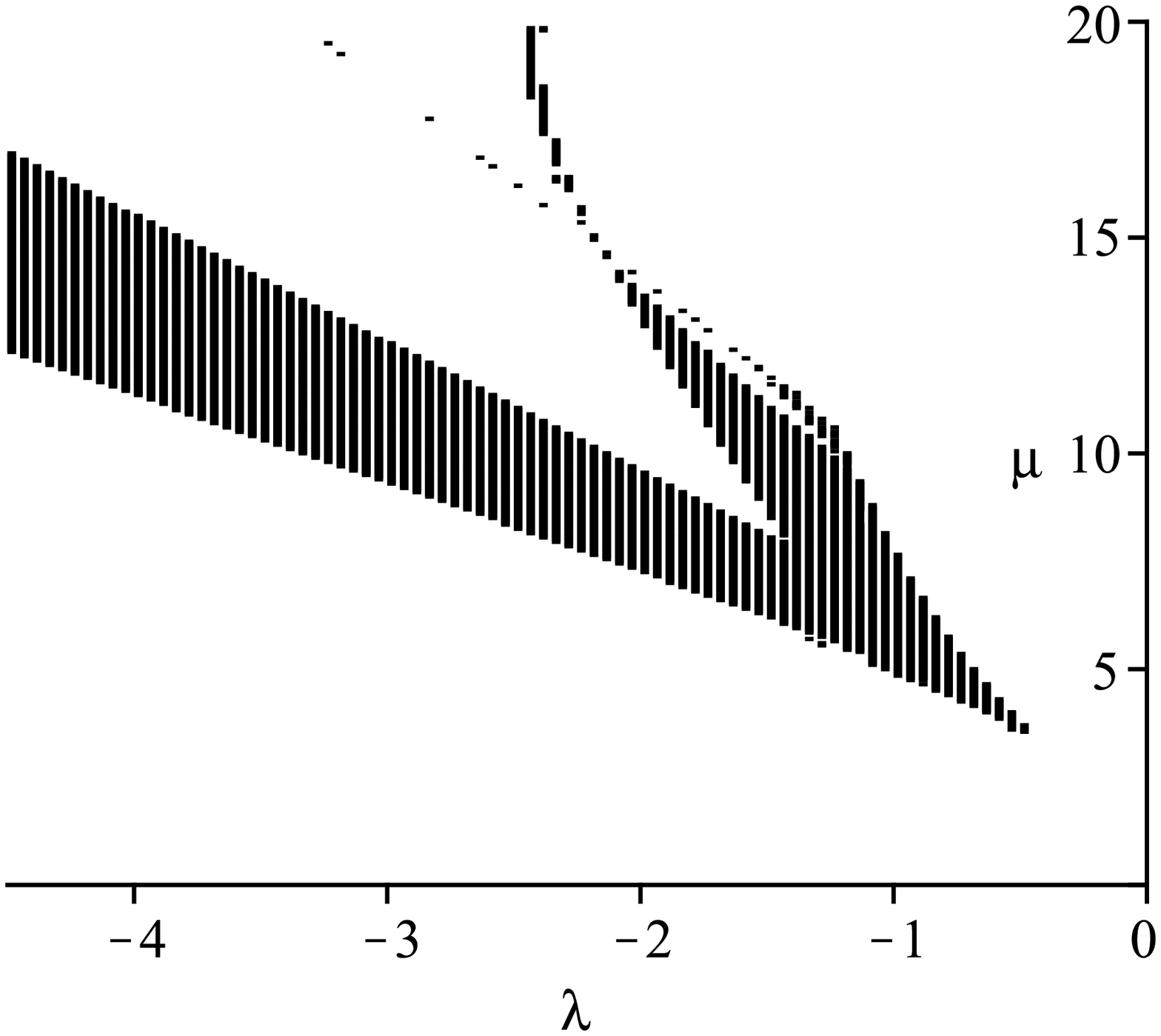} \includegraphics[scale=0.25]{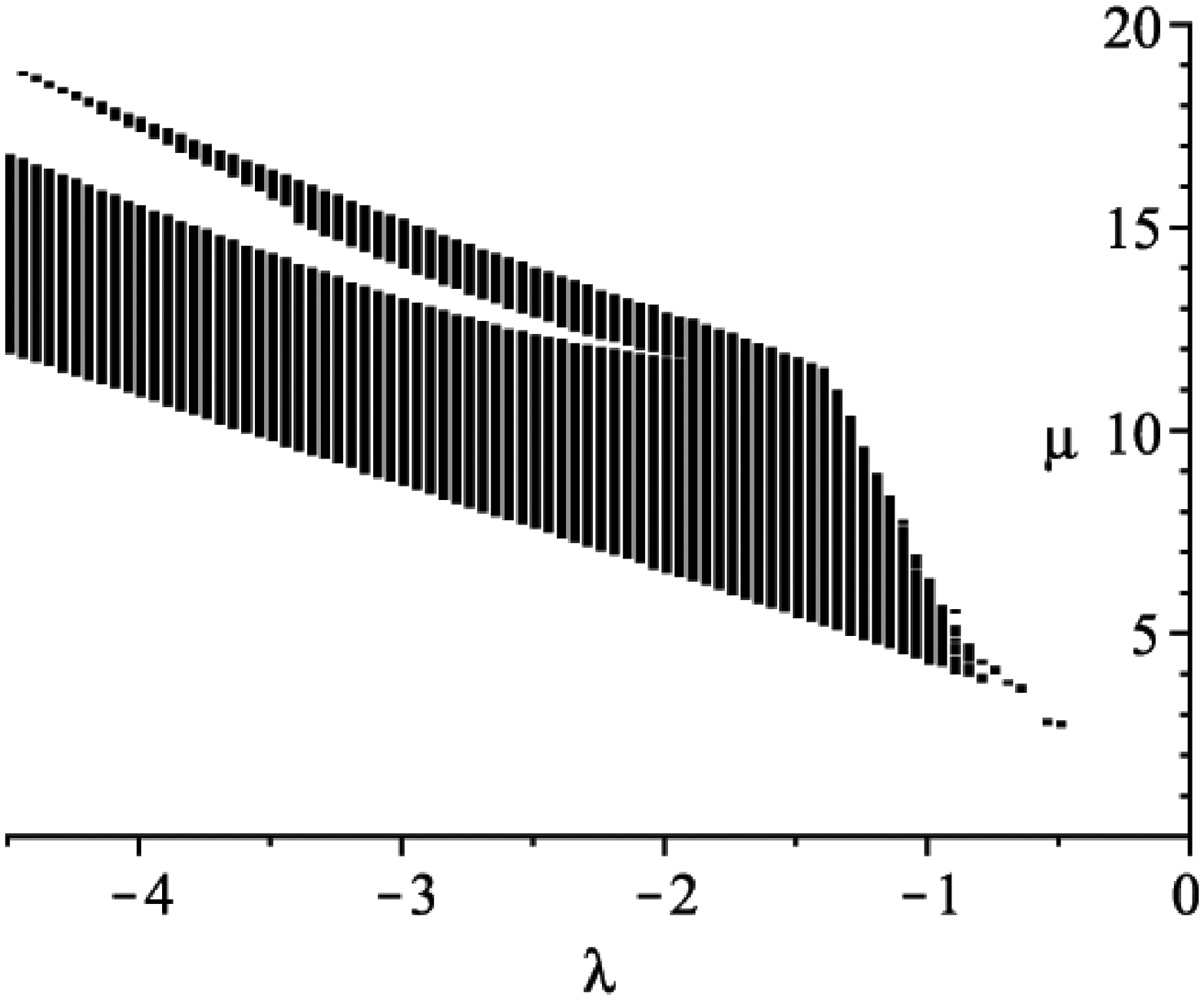}
\vspace{-0.2in}
\begin{equation}
\text{(a)} \qquad \qquad \qquad \qquad \qquad \qquad \text{(b)}\notag
\end{equation}
\includegraphics[scale=0.25]{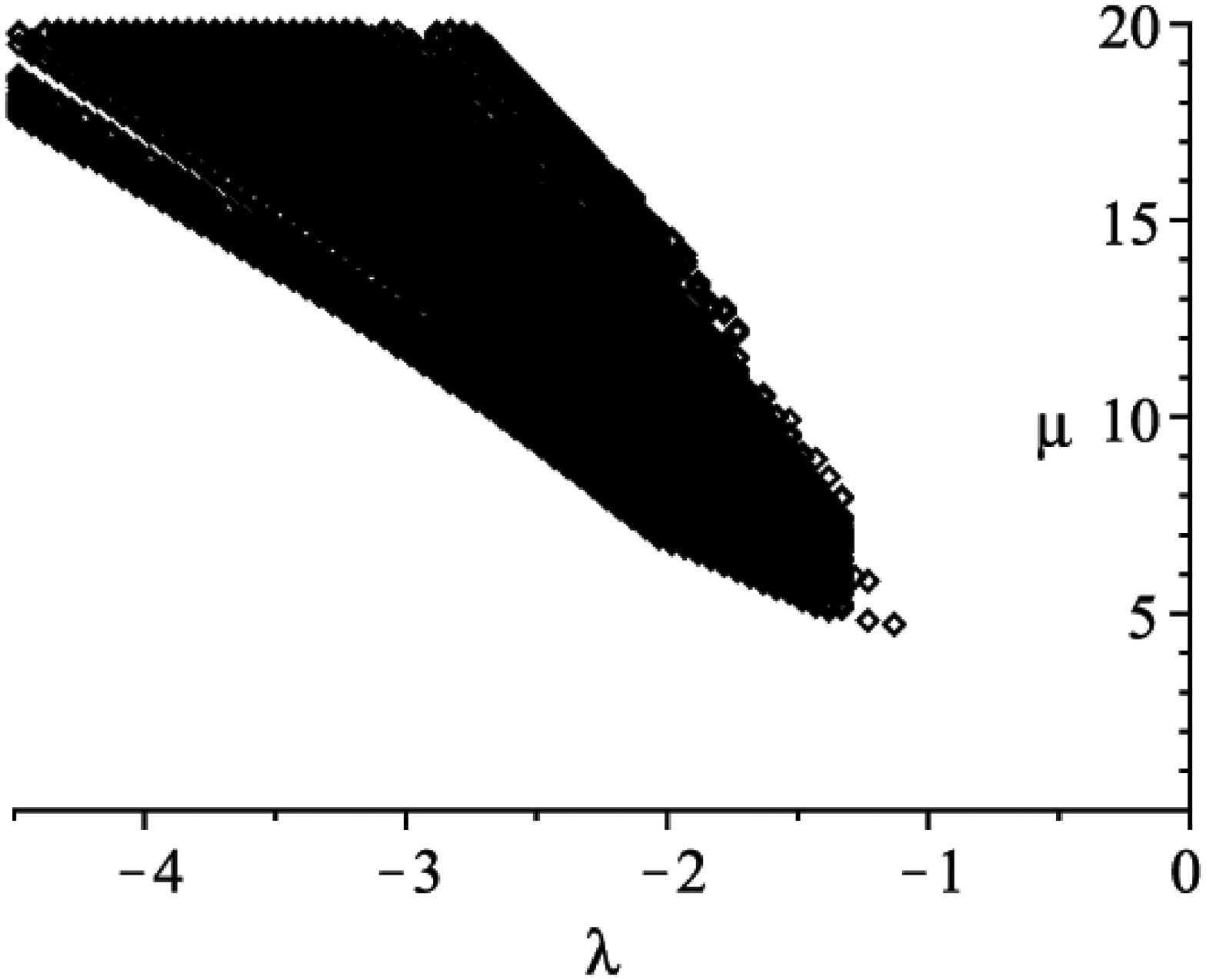} \includegraphics[scale=0.25]{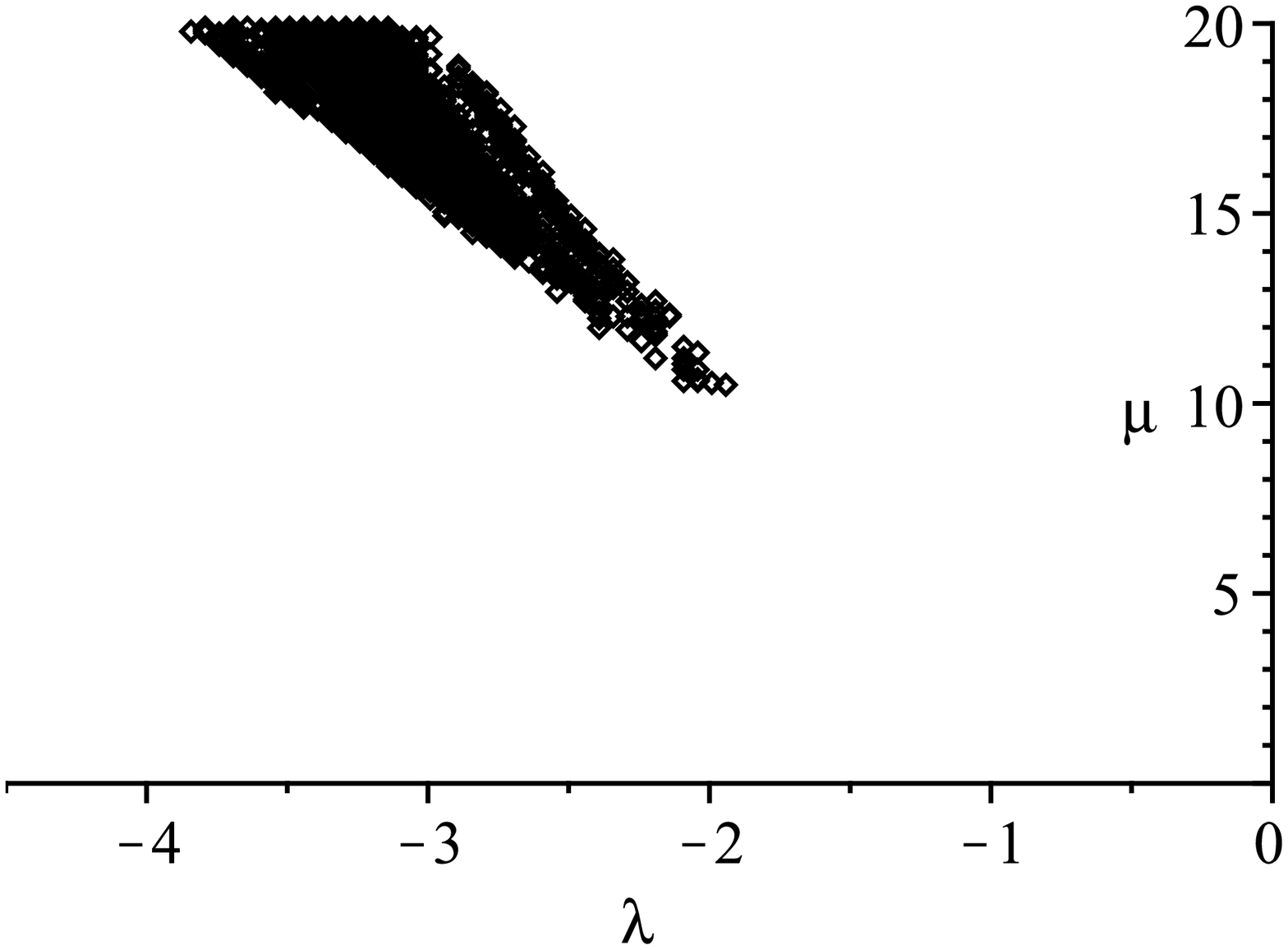}
\vspace{-0.2in}
\begin{equation}
\text{(c)} \qquad \qquad \qquad \qquad \qquad \qquad \text{(d)}\notag
\end{equation}
\vspace{-0.5in}
\end{center}
\caption{The same as in Fig. 1 [except that the blue line (\protect\ref{ml})
is not shown] for $\protect\beta =-2$ and (a) $m=1$, (b) $m=2$, (c) $m=5$,
(d) $m=10$.}
\end{figure}

Examples of the solitons, including the juxtaposition of their numerically
found and VA-predicted profiles, are displayed in Fig. 3 for $\beta =-0.5$
and $m=1$. Fixing $\lambda =-4$, we pick solutions from the larger lower
stability region and the upper one in Fig. 1(a) corresponding to $\mu =2.6$
and $\mu =3.55$, respectively. Note that, while the profiles of the
unstaggered component $u_{n}$ are very similar to one another, the solutions
for $v_{n}$ are different. In both cases, the variational solutions agree
well with their numerical counterparts.
\begin{figure}[tbp]
\begin{center}
\includegraphics[scale=0.35]{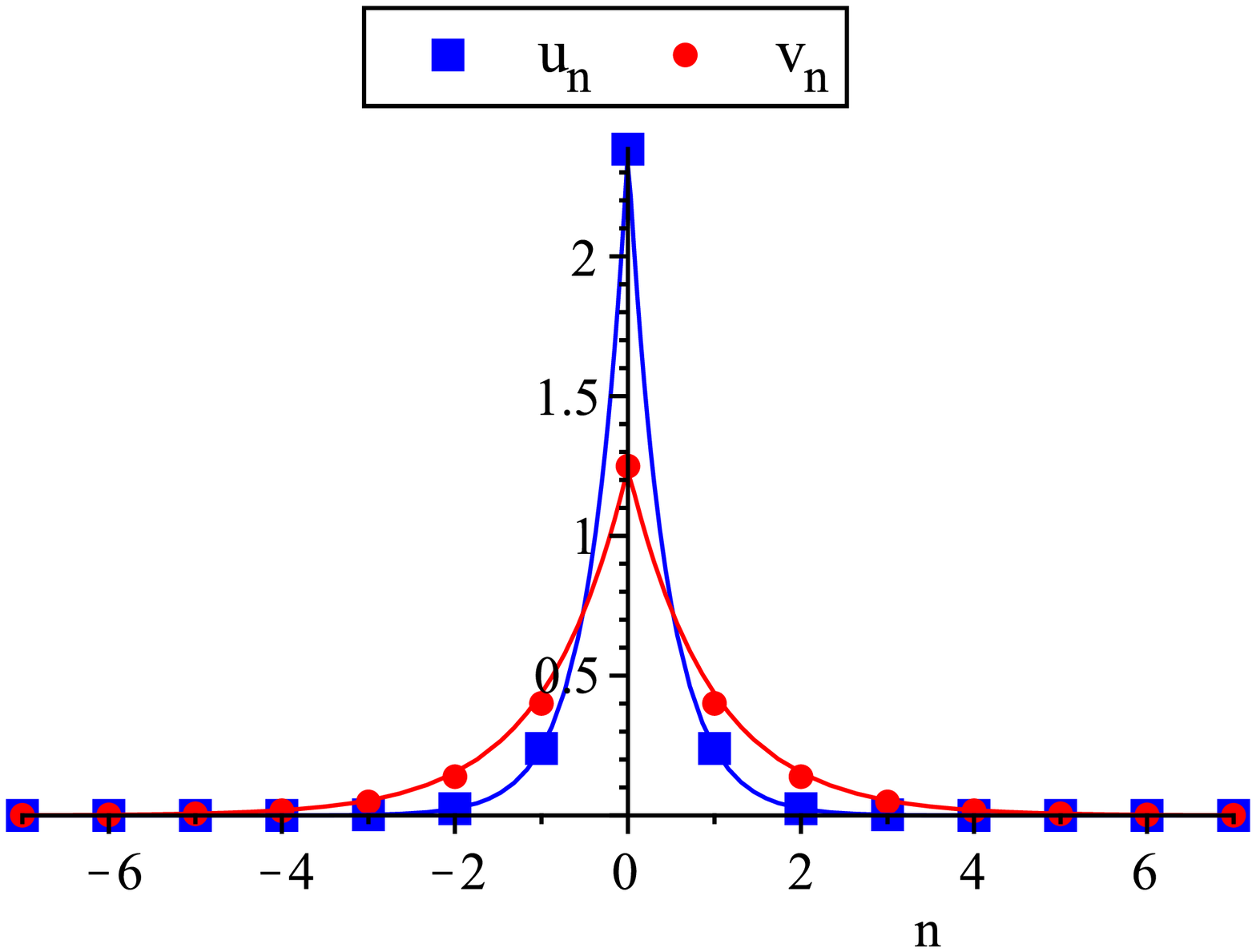} %
\includegraphics[scale=0.35]{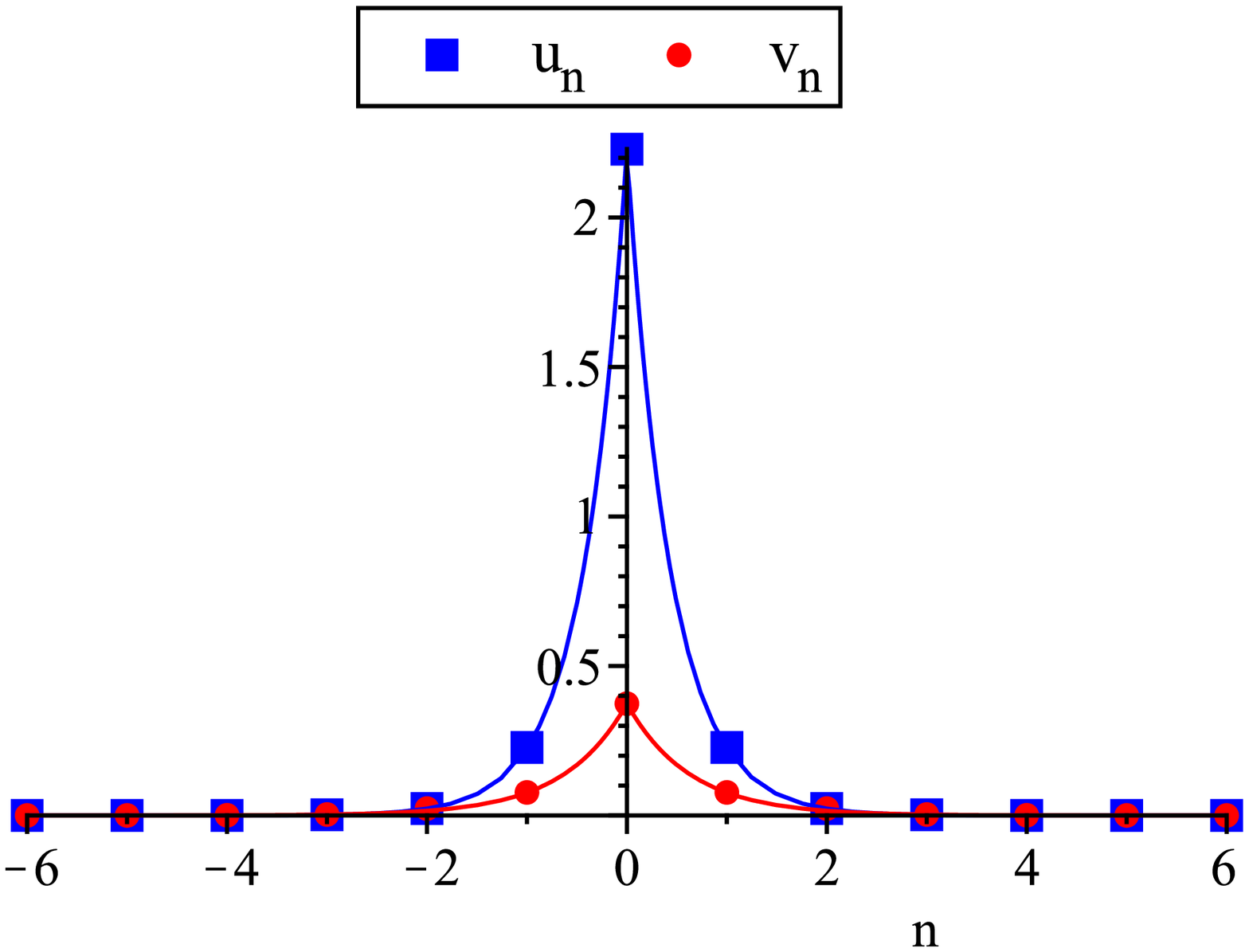} \vspace{-0.2in}
\begin{equation}
\text{(a)} \qquad \qquad \qquad \qquad \qquad \qquad \text{(b)}\notag
\end{equation}
\vspace{-0.5in}
\end{center}
\caption{(Color online) Examples of discrete solitons for $\protect\beta %
=-0.5$, $m=1$ and $(\protect\lambda ,\protect\mu )=(-4,2.6)$ (a), or $(%
\protect\lambda ,\protect\mu )=(-4,3.55)$ (b). Symbols and lines depict the
numerical solutions and prediction of the variational approximation,
respectively. Both solitons shown here are stable.}
\end{figure}

In Fig. 4, the solitons are plotted for the four branches of the existence
region in Fig. 1(c) when $\beta =-2$ and $m=1$. The first solution, shown in
Fig. 4(a), belongs to a very narrow existence branch, which is barely
discernible in Fig. 1(c) (its vertical width is $\Delta \mu <0.002$), and
exists along the bottom right of the main existence region (near the edge 
where $\lambda \approx -1.3$ and $\mu \approx 5.0$). The other
solutions are taken from the large lower existence region [Fig. 4(b)], the
large upper one [Fig. 4(c)], and the thin upper stripe which splits off from
the large upper branch [Fig. 4(d)]. Solutions from the lowest region [Fig.
4(a)] feature wider profiles in $u_{n}$ (note that both $u_{\pm 1}$ for them
are on the same order of magnitude at $u_{0}$) than do the solutions from
all the other branches, which exhibit sharp profiles and agree well with the
VA. On the contrary, the broad profile for $u_{n}$ in Fig. 4(a) cannot be
approximated properly by the exponential ansatz (\ref{ansatz}).
\begin{figure}[tbp]
\begin{center}
\includegraphics[scale=0.35]{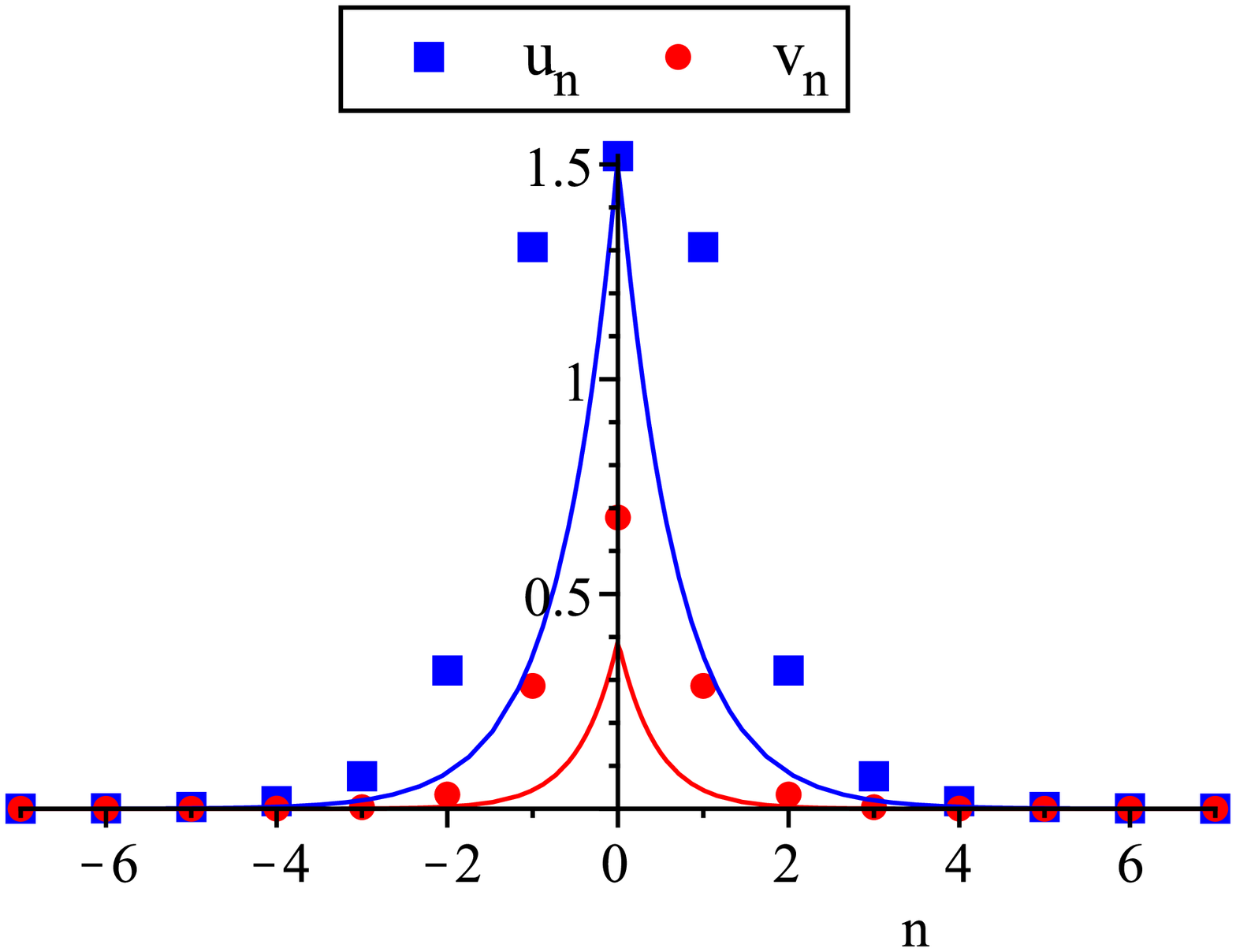} %
\includegraphics[scale=0.35]{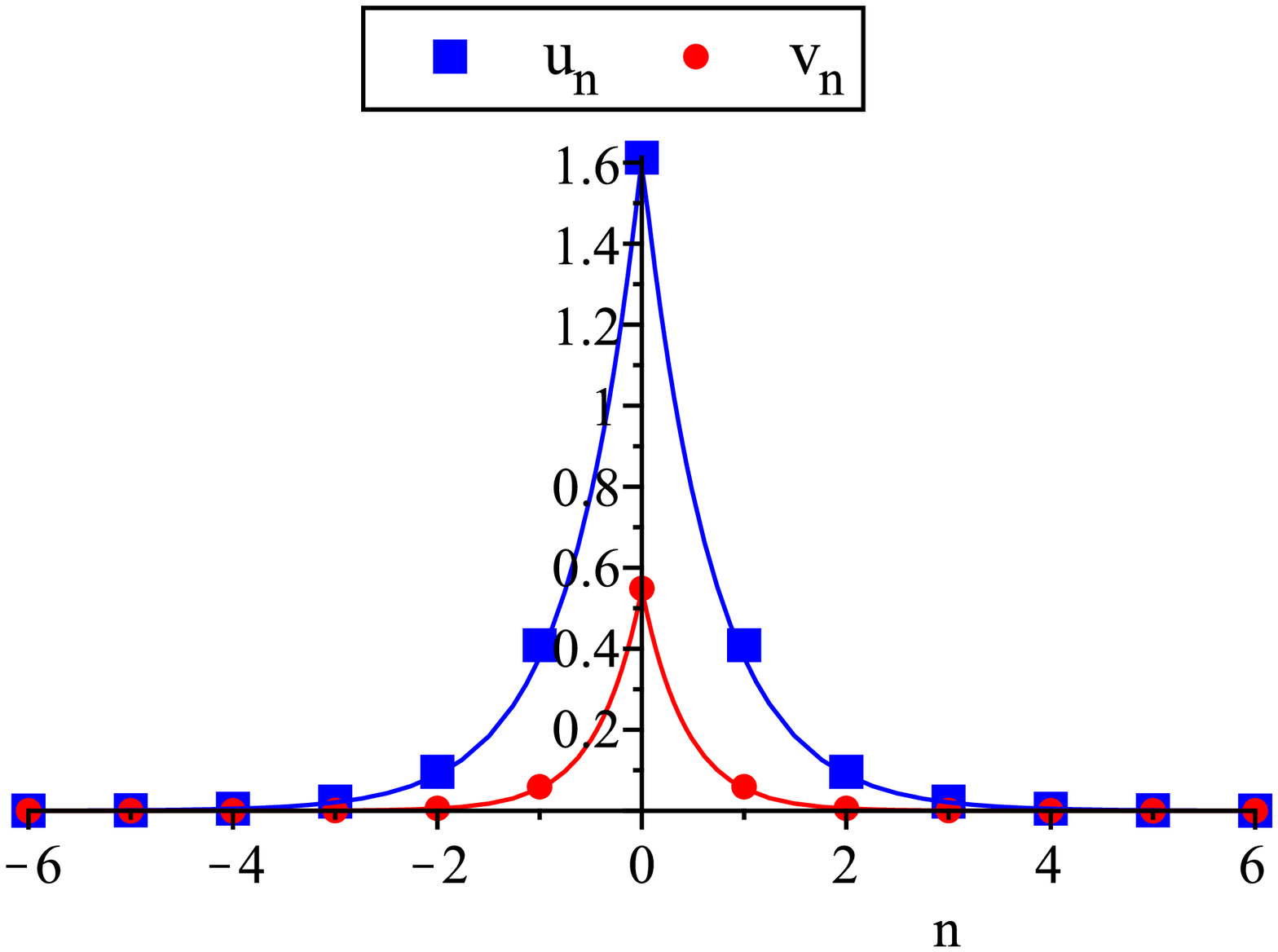} \vspace{-0.2in}
\begin{equation}
\text{(a)} \qquad \qquad \qquad \qquad \qquad \qquad \qquad \qquad \text{(b)}\notag
\end{equation}
\includegraphics[scale=0.35]{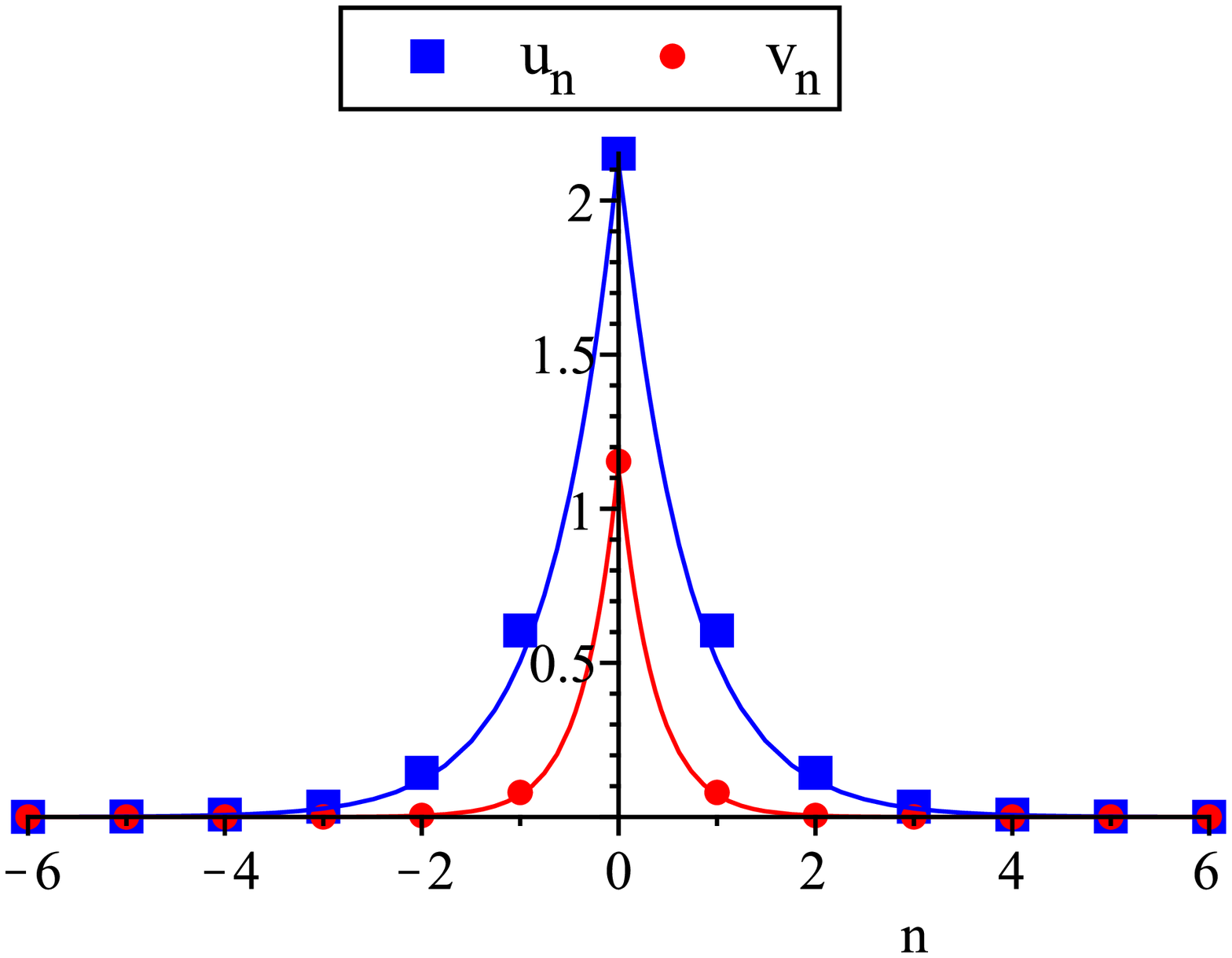} %
\includegraphics[scale=0.35]{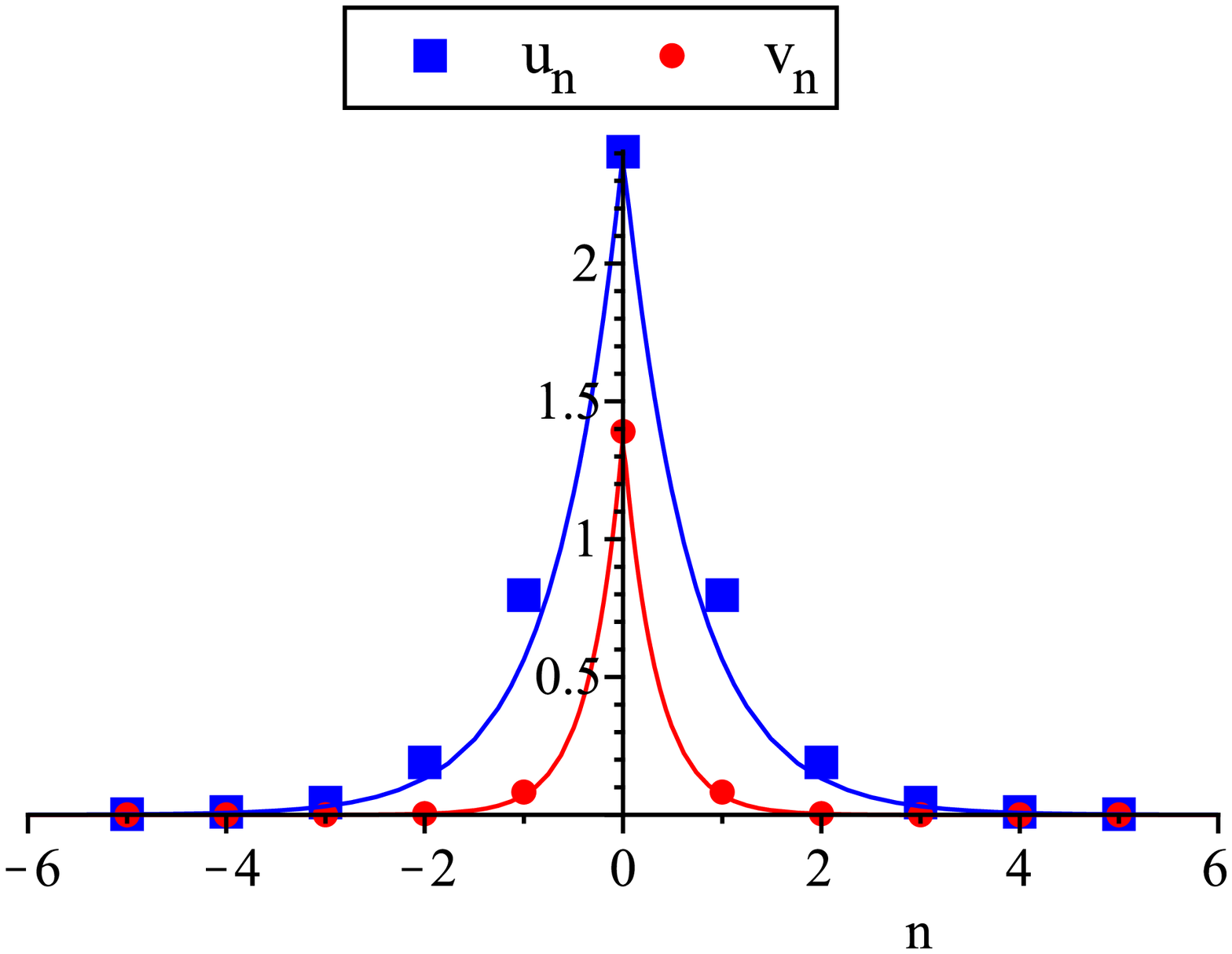} \vspace{-0.2in}
\begin{equation}
\text{(c)} \qquad \qquad \qquad \qquad \qquad \qquad \qquad \qquad \text{(d)}\notag
\end{equation}
\vspace{-0.5in}
\end{center}
\caption{(Color online) The same as in Fig. 3, for $\protect\beta =-2$, $m=1$
and $(\protect\lambda ,\protect\mu )=(-1.25,5.58)$ (a), $(\protect\lambda ,%
\protect\mu )=(-1.25,6.005)$ (b), $(\protect\lambda ,\protect\mu )=(-1.25,9)$
(c), and $(\protect\lambda ,\protect\mu )=(-1.25,10.7)$ (d). For the soliton
in panel (a), the variational approximation provides a poor fit to the
numerical solution, and this soliton is unstable. Other solitons are well
approximated by the variational ansatz, and are stable.}
\end{figure}

Because the numerical method employed here starts in a region where
the variational equations, Eqs. \eqref{eqs}, have a solution, we
cannot be absolutely sure that numerical solutions exist only in the
dark areas shown in Figs. 1 and 2. In principle, other branches of
numerical solutions might exist too, being unrelated to the VA,
although this does not seem plausible.

\subsection{Soliton stability}

Systematic simulations of the evolution of perturbed solitons shown
in Fig. 3 and Fig. 4 confirm that they are stable, with the
exception of the one in Fig. 4(a). Further, systematic tests clearly
suggest that the numerically found solitons are stable if their
shapes are close to those predicted by the VA, whereas ``broad"
solutions, which disagree with the VA, turn out to be unstable.
Actually, such unstable solitons are found only near the lower
boundary of the regions shown in Fig. 1.

In order to deduce the stability in a more general way, we define the
energies (norms) of the components,
\begin{equation}
W_{u}(\lambda ,\mu )=\sum_{n=-\infty }^{\infty }|u_{n}|^{2}\,,~W_{v}(\lambda
,\mu )=\sum_{n=-\infty }^{\infty }|v_{n}|^{2}\,.  \label{Wu}
\end{equation}%
In Fig. 5 $W_{u}(\lambda ,\mu )$ and $W_{v}(\lambda ,\mu )$ are plotted for $%
m=1$ and $\beta =0.5$, and in Figs. 6 and 7 we do the same for $\beta =-1.1$
and $\beta =-2$, respectively. There is a noticeable difference between the
energy surfaces for the $\beta <-1$ and $-1<\beta <0$ cases. In all cases
considered, the energy surfaces are monotonous in $\lambda $ and $\mu $ over
the stability regions (this finding agrees with the stability results
reported in Ref. \cite{MW1996}). However, for $-1<\beta <0$, the energy
surfaces increase in $\lambda $ (holding $\mu $ fixed) and decrease in $\mu $
(holding $\lambda $ fixed), as seen in Fig. 5. The opposite feature is
observed at $\beta <-1$: the energy surfaces decrease in $\lambda $ at fixed
$\mu $, and increase in $\mu $ at fixed $\lambda $.
\begin{figure}[tbp]
\begin{center}
\includegraphics[scale=0.6]{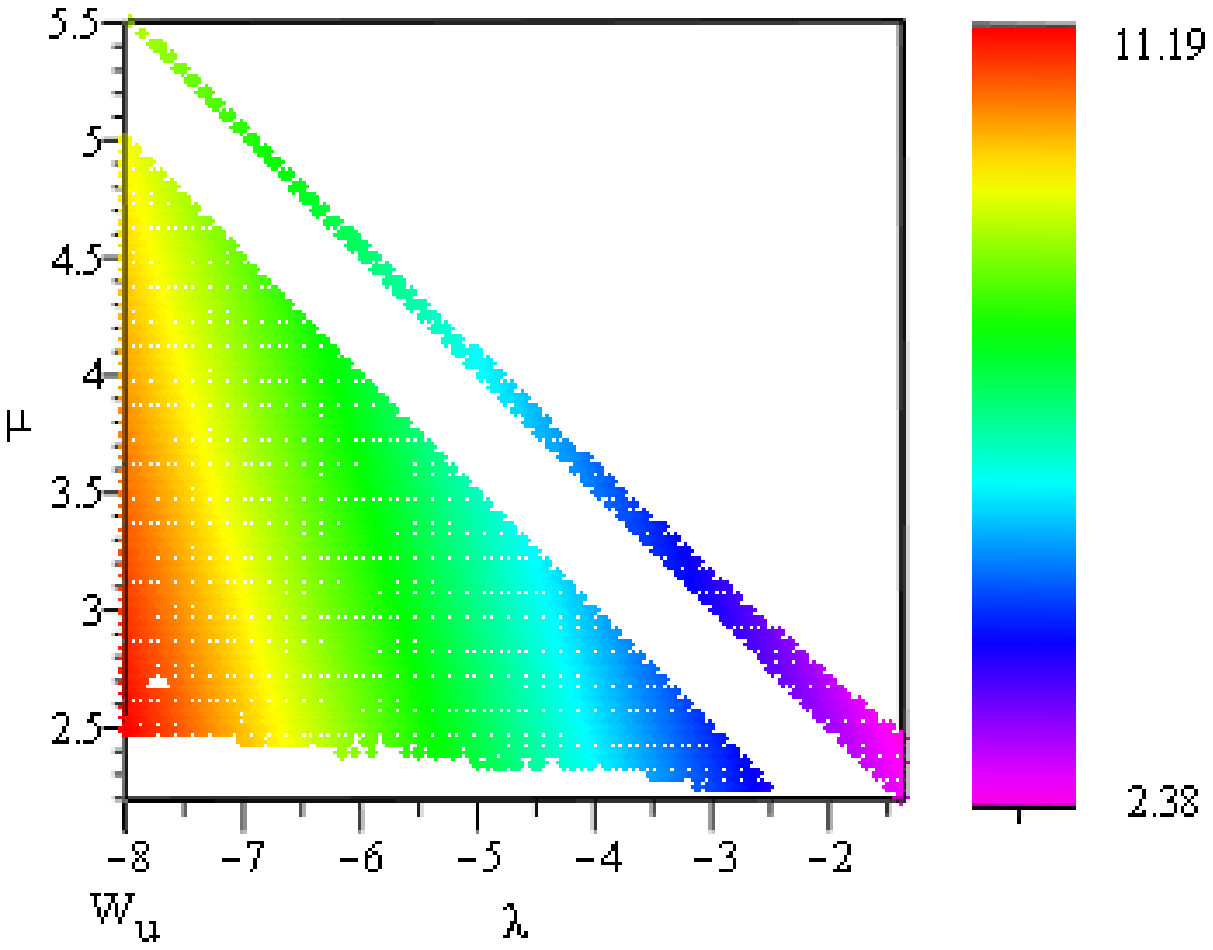} \includegraphics[scale=0.6]{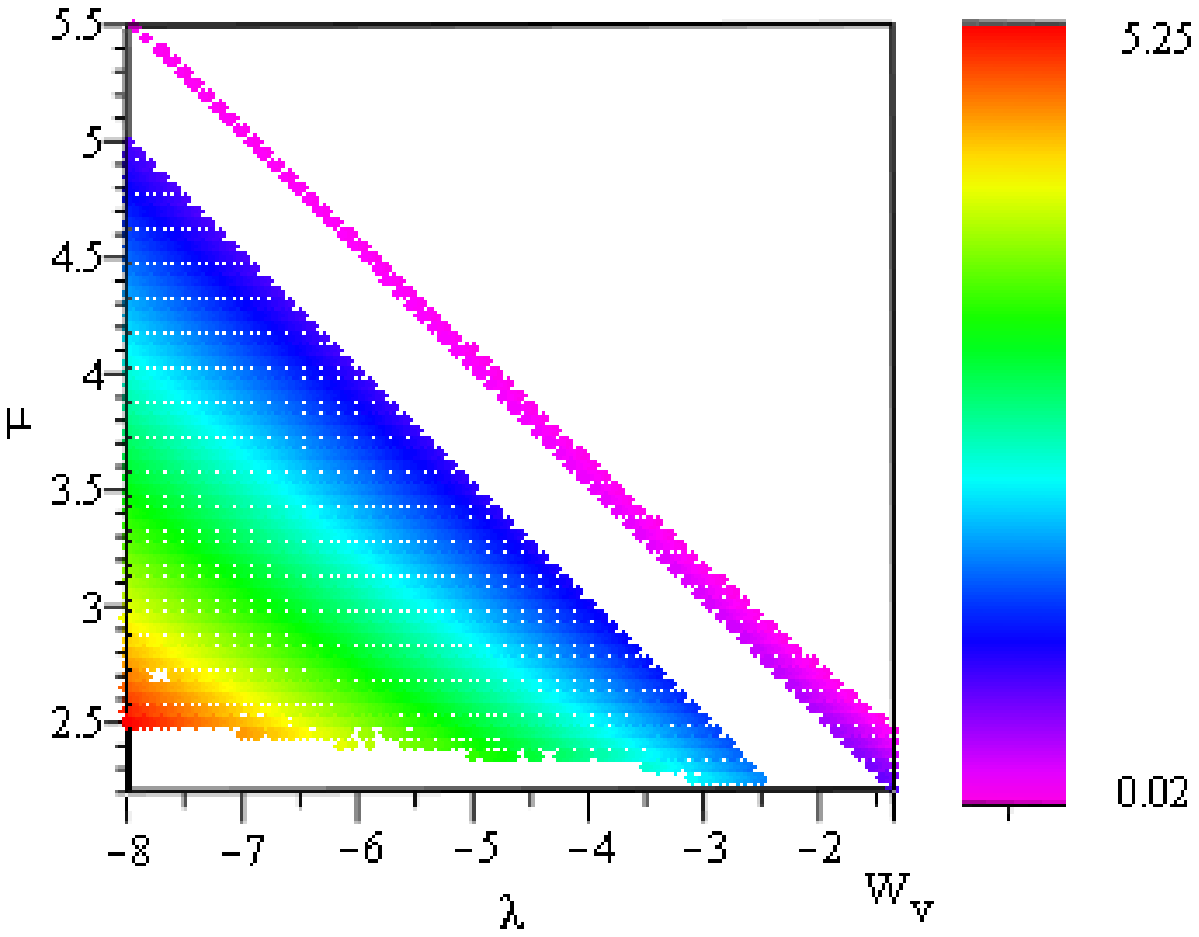}
\begin{equation}
\text{(a)} \qquad \qquad \qquad \qquad \qquad \qquad \qquad \qquad \text{(b)}\notag
\end{equation}
\vspace{-0.5in}
\end{center}
\caption{(Color online) Color-coded plots of the energy surfaces $W_{u}(%
\protect\lambda ,\protect\mu )$ (a) and $W_{v}(\protect\lambda ,\protect\mu )
$ (b) in the existence regions for the discrete solitons in the $(\protect%
\lambda ,\protect\mu )$ plane, at $m=1$ and $\protect\beta =-0.5$. The
energy increases in $\protect\lambda $ (holding $\protect\mu $ fixed) and
decreases in $\protect\mu $ (holding $\protect\lambda $ fixed).}
\end{figure}
\begin{figure}[tbp]
\begin{center}
\includegraphics[scale=0.6]{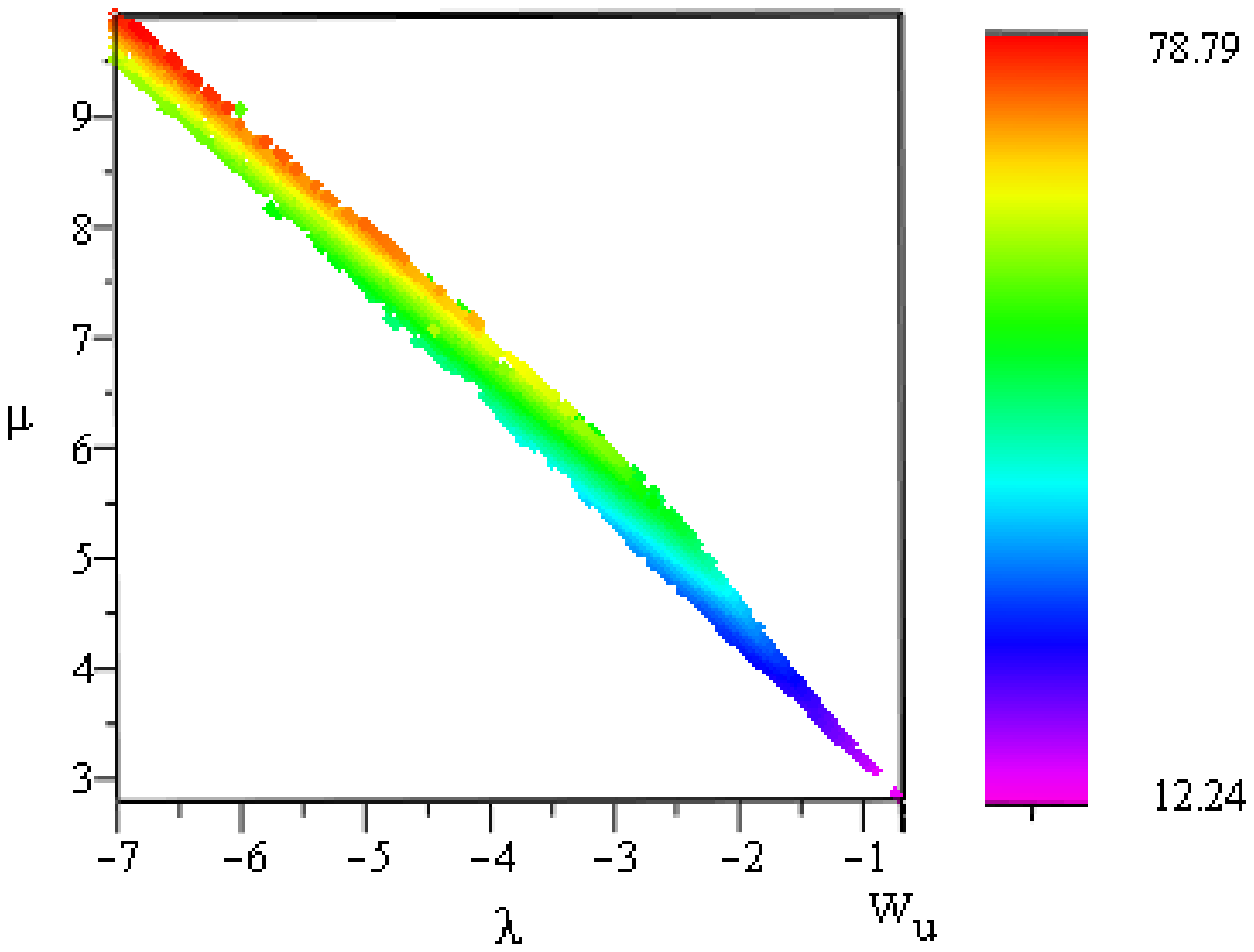} \includegraphics[scale=0.6]{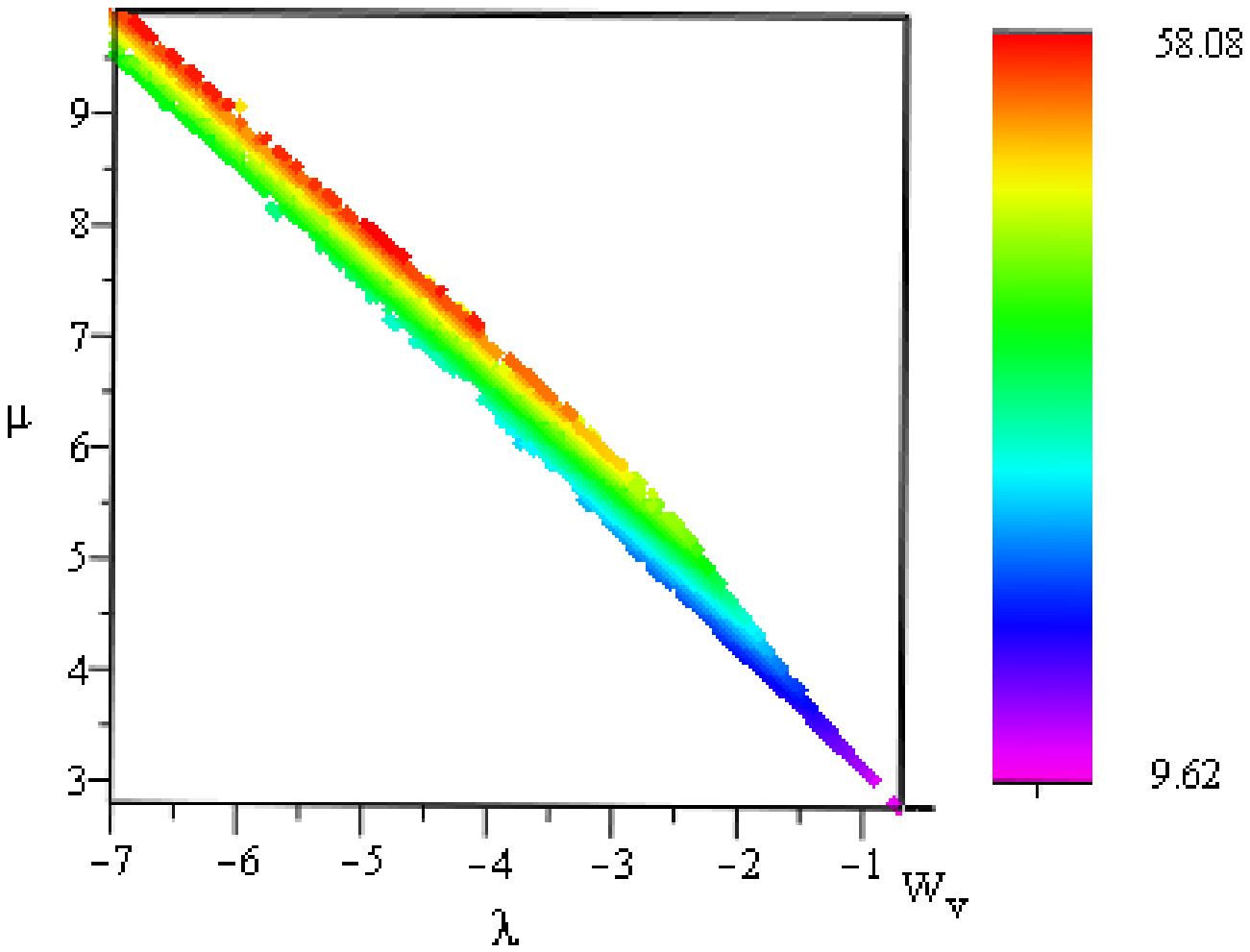}
\begin{equation}
\text{(a)} \qquad \qquad \qquad \qquad \qquad \qquad \qquad \qquad \text{(b)}\notag
\end{equation}
\vspace{-0.5in}
\end{center}
\caption{(Color online) The same as in Fig. 5, but for $m=1$ and $\protect%
\beta =-1.01$. The energy surface decreases in $\protect\lambda $ (holding $%
\protect\mu $ fixed) and increase in $\protect\mu $ (holding $\protect%
\lambda $ fixed).}
\end{figure}
\begin{figure}[tbp]
\begin{center}
\includegraphics[scale=0.6]{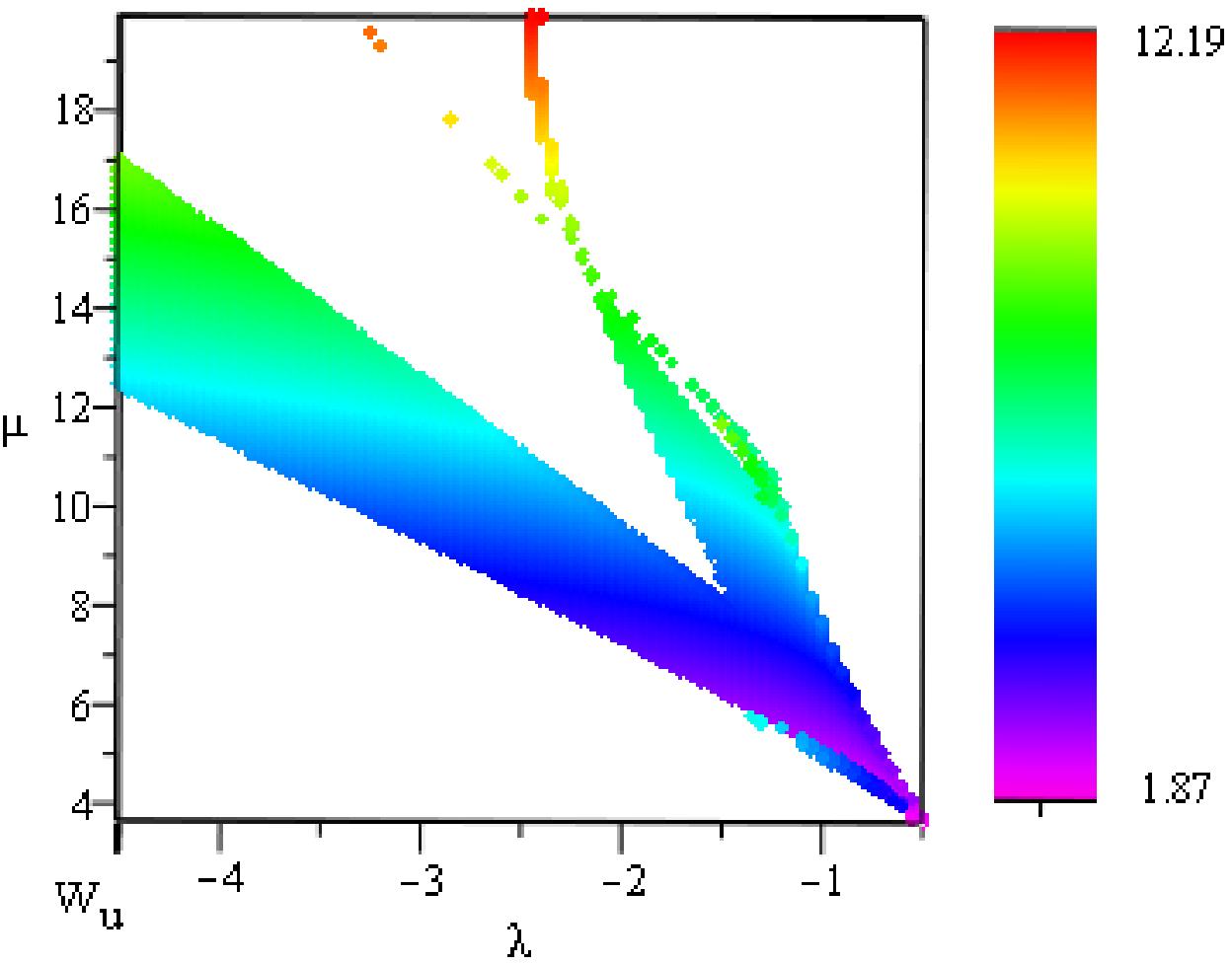} \includegraphics[scale=0.6]{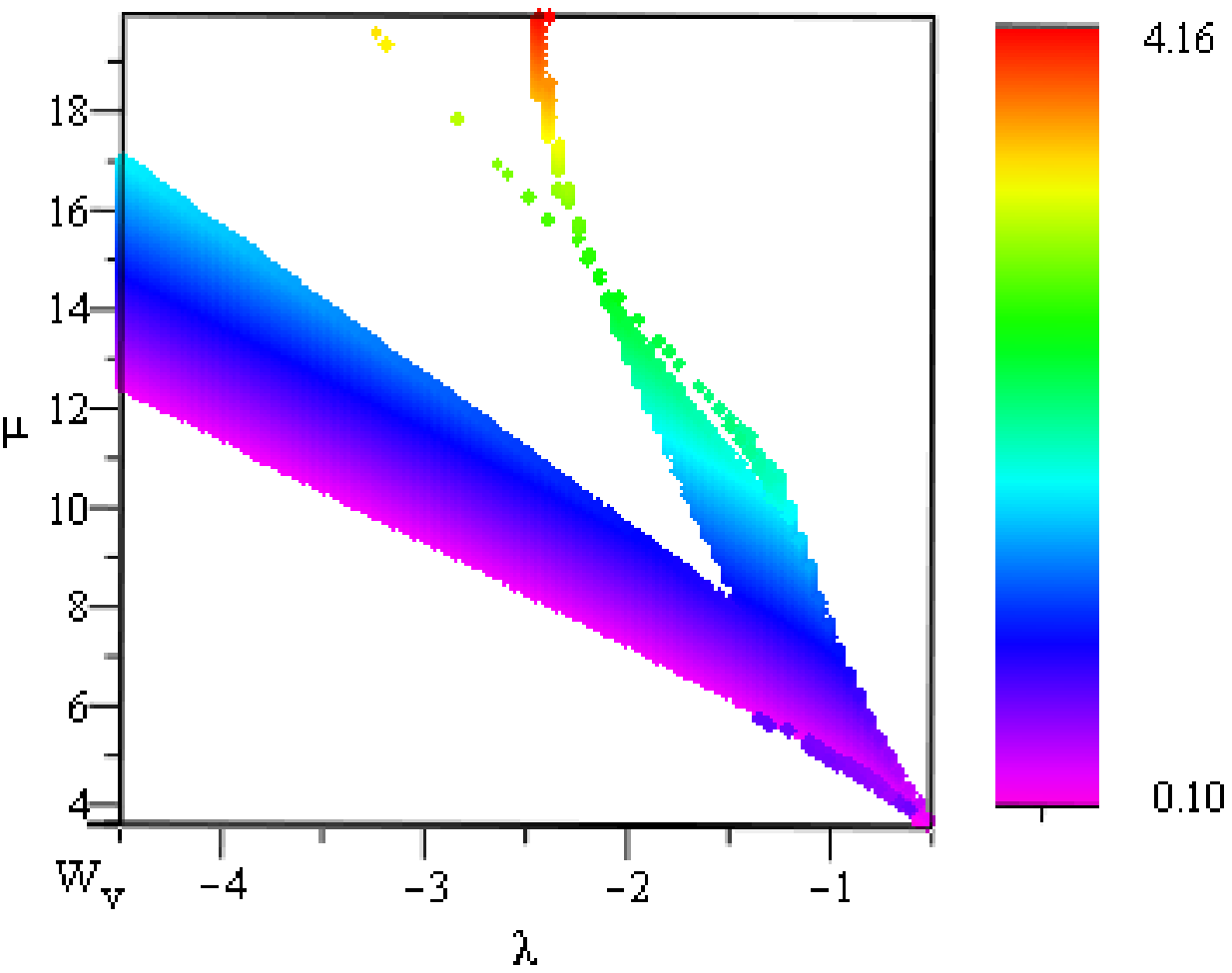}
\begin{equation}
\text{(a)} \qquad \qquad \qquad \qquad \qquad \qquad \qquad \qquad \text{(b)}\notag
\end{equation}
\vspace{-0.5in}
\end{center}
\caption{(Color online) The same as in Figs. 5 and 6, but for $m=1$ and $%
\protect\beta =-2$. The energy decreases in $\protect\lambda $ (holding $%
\protect\mu $ fixed) and increases in $\protect\mu $ (holding $\protect%
\lambda $ fixed).}
\end{figure}

More can be stated about the stability by means of the VA. The substitution
of ansatz (\ref{ansatz}) into Eqs. \eqref{Wu} yields%
\begin{equation}
W_{u}(\lambda ,\mu )=A\sum_{n=-\infty }^{\infty }e^{-2p|n|}=A\coth (p)\,,
\end{equation}%
\begin{equation}
W_{v}(\lambda ,\mu )=B\sum_{n=-\infty }^{\infty }e^{-2q|n|}=B\coth (q)\,.
\end{equation}%
where $A$ and $B$ are functions of $\lambda $ and $\mu $ determined by Eqs. (%
\eqref{eqs}). As is known from the generalized VK criterion for systems with
two conserved norms \cite{Berge'}, a stability change occurs when Jacobian $%
\partial (W_{u},W_{v})/\partial (\lambda ,\mu )$ changes its sign. We follow
this approach in Fig. 8, where the zero locus of the Jacobian is plotted,
along with the region of the existence of the numerically found solitons,
for $\beta =-2$ and $m=1$. It is observed that the stability change
predicted by the VA nearly coincides with the lower boundary of the
existence region. The agreement is not perfect since the VA does not produce
exact results, but the mismatch is quite small. The majority of the soliton
solutions, which are located above the stability-change locus, are stable;
unstable are the solitons, such as the broad one displayed in Fig. 4(a),
which are found in a tiny area adjacent to the lower boundary which is
actually bounded by the Jacobian's zero locus crossing the existence region.
\begin{figure}[tbp]
\begin{center}
\includegraphics[scale=0.4]{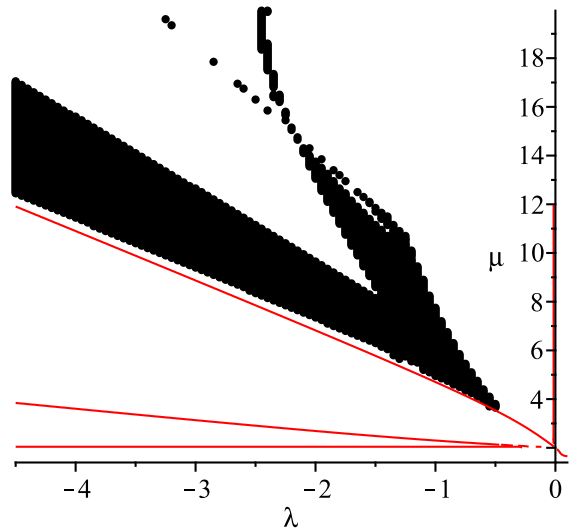} \vspace{-0.3in}
\end{center}
\caption{(Color online) The numerically found existence regions (black) in
the $(\protect\lambda ,\protect\mu )$ plane for the single-peak solitons,
juxtaposed with the (red online / dark grey in print) curves defined by $\partial (W_{u},W_{v})/\partial
(\protect\lambda ,\protect\mu )=0$, as produced by the variational
approximation, for $m=1$ and $\protect\beta =-2$. The solitons existing
above the red line are stable.}
\end{figure}

Comparing Figs. 5-7, we see that as $\beta \rightarrow -1$, the energy
values for the obtained solitons increase. We observe that this, in turn,
corresponds to a change in the stability, and it was found that solitons
occurring in the narrow region shown in Fig. 1(b) for $\beta =-1.01$ are
unstable. One such soliton is plotted in Fig. 9. Notice that the unstable
numerical solution is again essentially wider than its variational
counterpart, i.e., as in Fig. 4(a), the variational approximation is a poor
fit to the broad soliton. We have tested the stability of similar soliton
solutions for $\beta =-1.10$ and $\mu =6.0$, and have found them to be
stable. So the instability region appears to be localized around $\beta =-1$%
.
\begin{figure}[tbp]
\begin{center}
\includegraphics[scale=0.35]{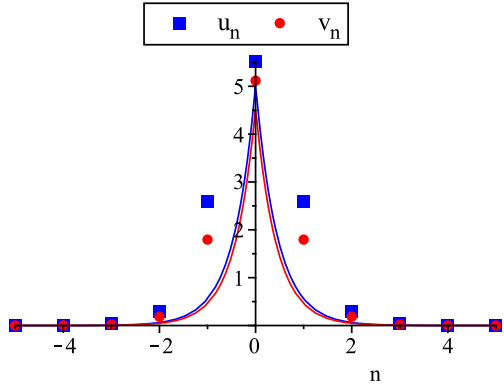} \vspace{-0.35in}
\end{center}
\caption{(Color online) An example of a discrete soliton for $\protect\beta %
=-1.01$, $m=1$ and $(\protect\lambda ,\protect\mu )=(-3.55,6)$. Symbols and
lines depict the numerical solutions and prediction of the variational
approximation, respectively. The soliton shown here is unstable.}
\end{figure}
Lastly, direct simulations demonstrate that the unstable broad soliton
solutions, such as the one displayed in Fig. 4(a), decay into a combination
of multiple breathers and emitted radiation, as seen in the example in Fig. 10.
\begin{figure}[tbp]
\begin{center}
\includegraphics[scale=0.4]{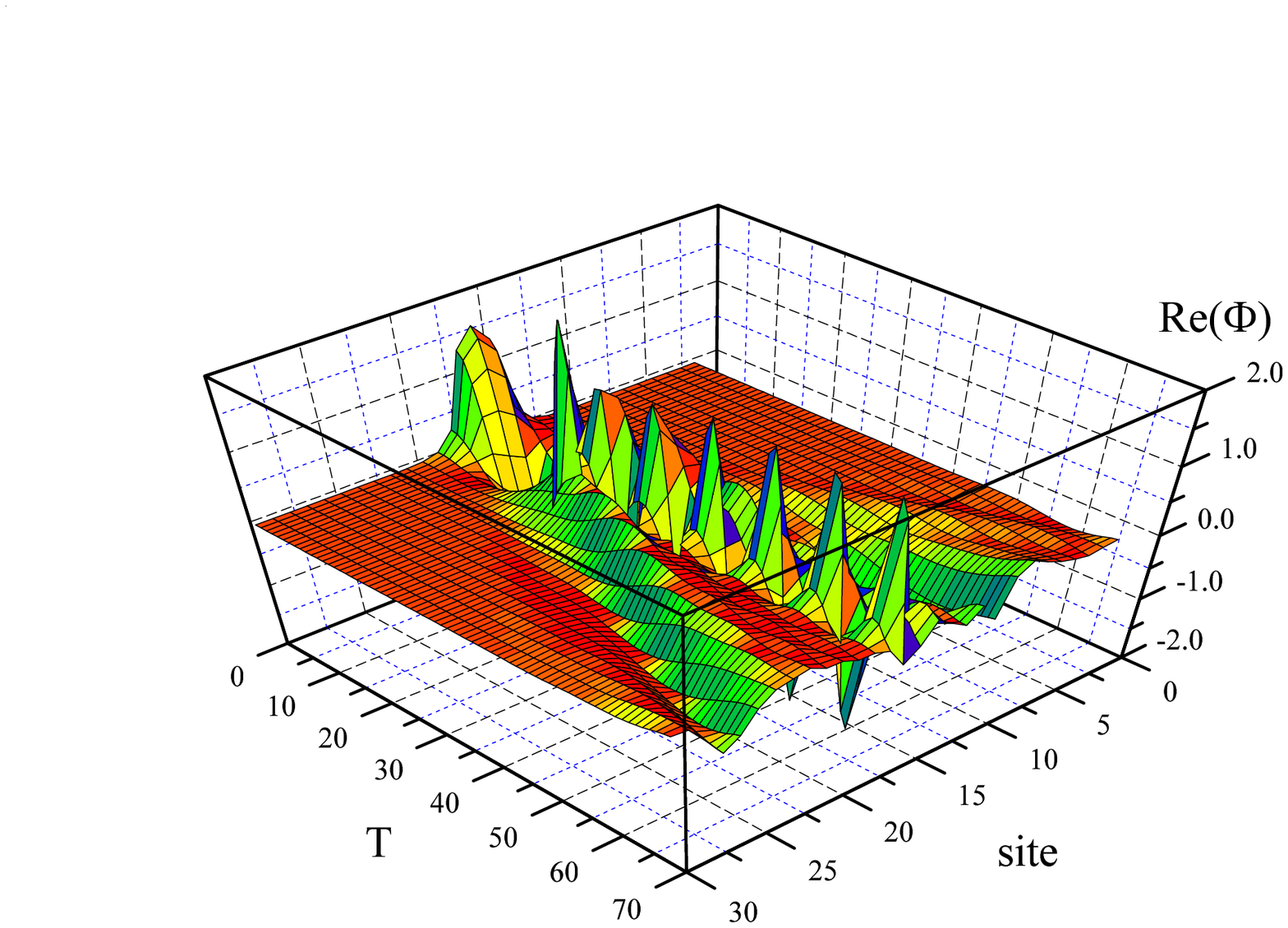} \includegraphics[scale=0.4]{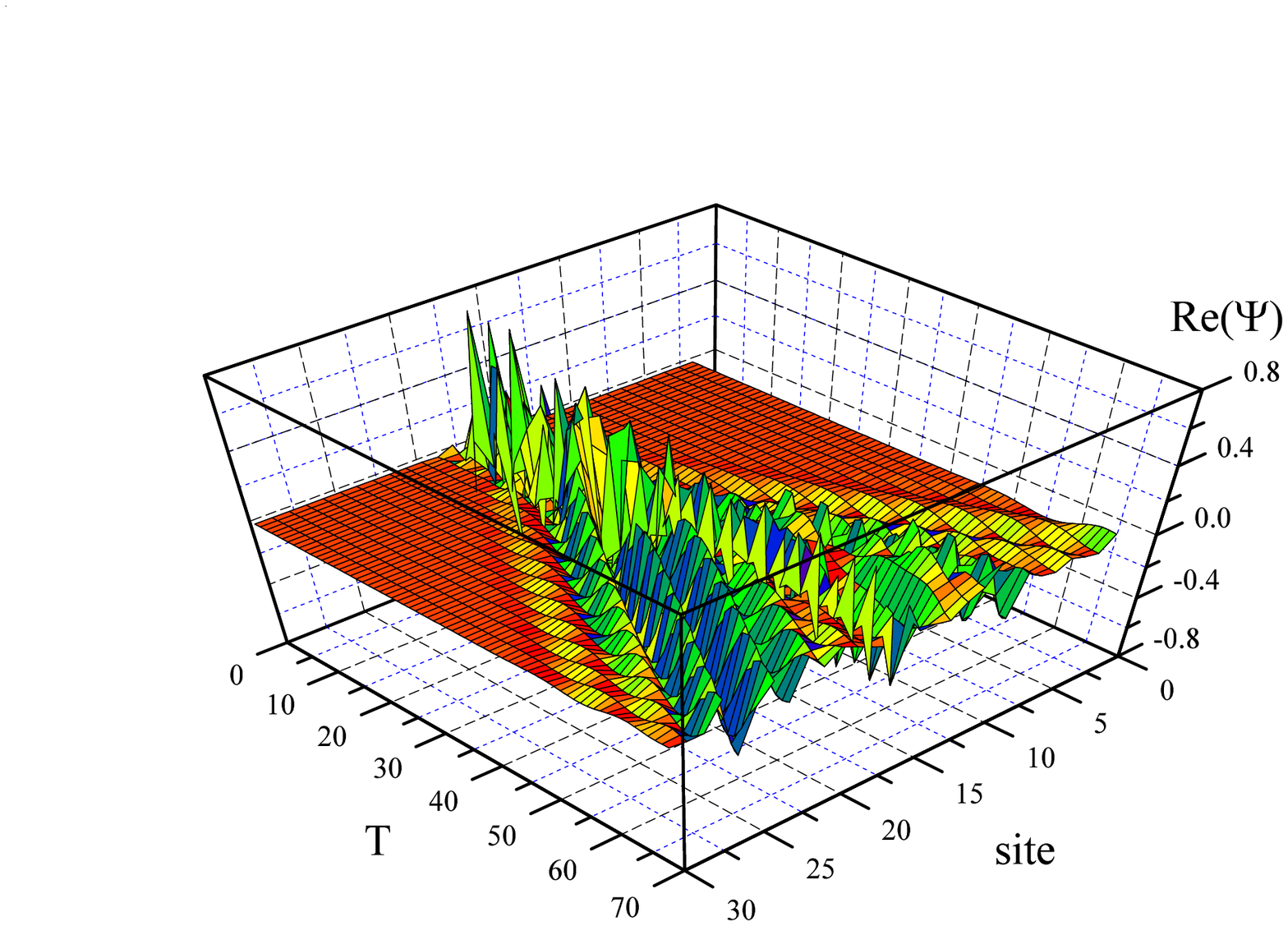}
\vspace{-0.2in}
\begin{equation}
\text{(a)} \qquad \qquad \qquad \qquad \qquad \qquad \qquad \qquad \text{(b)}\notag
\end{equation}
\vspace{-0.5in}
\end{center}
\caption{(Color online) Plots of (a) $\protect\phi _{n}(t)$ and
(b) $\protect\psi _{n}(t)$ of the evolution of the unstable soliton from Fig. 4(a), which
corresponds to $\protect\beta =-2$, $m=1$, $(\protect\lambda ,\protect\mu %
)=(-1.25,5.58)$. Note the escaping radiation in both components, and the
oscillating breathers which are left behind.}
\end{figure}

\section{Conclusions}

We have introduced the symmetric system of DNLS (discrete nonlinear Schr\"{o}%
dinger) equations with the self-attractive on-site SPM nonlinearity and
repulsive XPM interaction, which supports two-component solitons of the
\textit{symbiotic} unstaggered-staggered type. The system may be implemented
in a mixture of two BEC species with identical or different atomic masses,
and (in principle) in arrays of bimodal optical waveguides. In the
analytical part of the work, \ the VA\ (variational approximation) was
developed, based on the exponential ansatz for the fundamental (single-peak)
solitons. In the limit of the large relative mass of the two species, the
TFA (Thomas-Fermi approximation) was elaborated too, which reduces the
coupled system to two different single-component DNLS equations in the inner
and outer layers of the solutions. Further, by means of the numerical
solution we have identified areas in the plane of the two chemical
potentials (propagation constants) where discrete solitons exist. It has
been inferred that the VA and TFA agree well with the numerical solutions,
except for a stripe near the lower existence boundary, where broad solitons
are poorly approximated by the exponential ansatz. Direct simulations of the
evolution of the perturbed solitons demonstrate that all the solitons which
are well approximated by the VA (i.e., almost all the solutions) are stable.
Only the broad solitons, which are not accommodated by the VA, are unstable.
The results for the stability can be accurately predicted by means of the
generalized VK criterion for the two-component system (with the stability
change corresponding to the vanishing of the respective Jacobian), realized
in terms of the VA.

It may be interesting to extend the work by considering multi-soliton
(multi-peak) bound states of the unstaggered-staggered type. A challenging
problem is to generalize the system for two-dimensional lattices and various
types of discrete two-dimensional solitons, including solitary vortices.

\section*{Acknowledgements}

B.A.M. appreciates a partial support from the Binational (US-Israel) Science
Foundation, through grant No. 2010239. R.A.V. was supported in part by a
National Science Foundation Graduate Research Fellowship.

%\clearpage

\end{document}